\documentclass[aps,prl,twocolumn,superscriptaddress,showpacs,floatfix,nobibnotes]{revtex4}
\usepackage{epsfig}
\usepackage{epstopdf}

\usepackage{graphicx}
\usepackage{longtable}
\usepackage{CJK}
\usepackage{color}

\usepackage{mathptmx, courier, pifont}
\usepackage[scaled=0.92]{helvet}
\usepackage[T1]{fontenc}
\usepackage{textcomp}

\begin{document}

\title{Triaxial rotor modes in finite-$N$ boson systems}
\author{Yu Zhang}
\affiliation{Department of Physics, Liaoning Normal University,
Dalian 116029, P. R. China}

\author{Sheng-Nan Wang}
\affiliation{Department of Physics, Liaoning Normal University,
Dalian 116029, P. R. China}

\author{Feng Pan}
\affiliation{Department of Physics, Liaoning Normal University,
Dalian 116029, P. R. China}\affiliation{Department of Physics and
Astronomy, Louisiana State University, Baton Rouge, LA 70803-4001,
USA}

\author{Chong Qi}
\affiliation{Department of Physics, KTH Royal Institute of
Technology, Stockholm 10691, Sweden}

\author{J. P. Draayer}
\affiliation{Department of Physics and Astronomy, Louisiana State
University, Baton Rouge, LA 70803-4001, USA}

\date{\today}

\begin{abstract}
We propose an algebraic approach to elucidate the dynamic characteristics of triaxial rotor modes in nuclei by mapping a triaxial rotor Hamiltonian to the interacting boson model (IBM) one within a finite-$N$ framework. Our method unveils striking features not observed in conventional modes, exemplified by the $B(E2)$ anomaly, characterized by $B(E2;4_1^+\rightarrow2_1^+)/B(E2;2_1^+\rightarrow0_1^+)<<1$. Using specific examples, we demonstrate that the peculiar properties of low-lying states in both neutron-rich and neutron-deficient Os nuclei can be comprehensively understood through the proposed Hamiltonian, which incorporates both rigid and soft triaxial rotor modes. This algebraic method not only offers fresh insights into triaxial dynamics but also showcases its capability in uncovering emergent exotic collective modes in nuclear structure.
\end{abstract}


\maketitle
\begin{center}
\vskip.2cm\textbf{1. Introduction}
\end{center}\vskip.2cm

The emergence of collective modes stands as a pivotal characteristic of complex nuclear many-body systems.
The systems with quadrupole deformation, in particular, exhibit several typical collective modes, such as the spherical vibrator, $\gamma$-unstable rotor, and deformed rotors (whether axial or triaxial). These collective modes represent the quantum mechanical expression of distinct intrinsic deformations inherent to nuclear systems \cite{Bohrbook}. The investigation of rotor modes in finite-$N$ systems has been a compelling pursuit since Elliott's groundbreaking work, which provided a microscopic elucidation of rotational spectra in light nuclei \cite{Elliott1958}.  Based on the algebraic relations \cite{Ui1970}, it was demonstrated that triaxial rotor dynamics can be microscopically elucidated through the SU(3) framework of the shell model \cite{Leschber1987,Castanos1988}. Similarly, the algebraic realization of rotor modes has been attained within the SU(3) limit of the interacting boson model (IBM) \cite{Smirnov2000}. An exact mapping scheme for the IBM realization \cite{Zhang2014}
 has been recently applied to explicate anomalous $E2$ transitions observed in neutron-deficient triaxial nuclei \cite{Zhang2022}. These endeavors underscore the intimate connection between rotor modes and SU(3) symmetry within nuclei \cite{Kotabook2020}.

The IBM~\cite{IachelloBook87} exhibits remarkable efficacy in elucidating collective modes within nuclei. Its notable advantage lies in its ability not only to encompass a few distinct and fundamental collective limits but, more crucially, to employ a simple Hamiltonian formulation~\cite{Warner1983} for investigating broader scenarios involving the mixing or transitions among various collective modes. However, the description of triaxial rotor modes, often employed in interpreting nuclear collective excitations~\cite{Davydov1958}, presents a challenge within the IBM due in part to the inability of the model to straightforwardly account for triaxial deformations via two-body terms. To address this limitation, one approach involves incorporating higher-order interactions~\cite{VC1981}. For instance, the inclusion of a cubic term $(d^\dag\times d^\dag\times d^\dag)^3\cdot(\tilde{d}\times \tilde{d}\times \tilde{d})^3$\cite{VC1981,Heyde1984} yields a potential surface with a stable triaxial minimum at $\gamma=30^\circ$, enhancing the IBM's description of triaxial nuclei\cite{Ramos2000I,Ramos2000II}. Additionally, it has been shown~\cite{Fortunato2011} that a parameter region of triaxiality within $0^\circ<\gamma<60^\circ$ can be accommodated in the extended consistent-$Q$ formalism, incorporating a cubic term in the quadrupole operators.
Broadly speaking, a scalar polynomial in the quadrupole operators can be used to generate
triaxial deformation within the IBM at the mean-field level \cite{Zhang2022,Teng2023}.
Notably, the rotor image within the IBM~\cite{Smirnov2000,Zhang2014} can be straightforwardly constructed using symmetry-conserving operators from the $\mathrm{SU(3)}\supset \mathrm{SO(3)}$ integrity basis, employing a scalar polynomial in the quadrupole operators \cite{Zhang2014,Zhang2022}. Despite achieving an algebraic realization within the SU(3) limit of the IBM, the impact of rotor modes on IBM dynamics remains to be fully elucidated. Particularly, a model analysis addressing the interplay between triaxial rotor and other collective modes is warranted, given that realistic nuclear systems often entail multiple collective modes.

In this work, we undertake a comprehensive examination of the IBM formulation of the triaxial rotor mode and its corresponding dynamics. Through this investigation, we aim to shed light on the pivotal role of the triaxial rotor in the IBM depiction of triaxial nuclei, particularly concerning anomalous $E2$ phenomena. Additionally, we endeavor to identify a unified framework capable of encompassing all typical collective modes, including the triaxial rotor, thus offering a cohesive description of nuclear dynamics.

\begin{center}
\vskip.2cm\textbf{2. Collective Modes in the IBM}
\end{center}\vskip.2cm

\begin{figure}
\begin{center}
\includegraphics[scale=0.15]{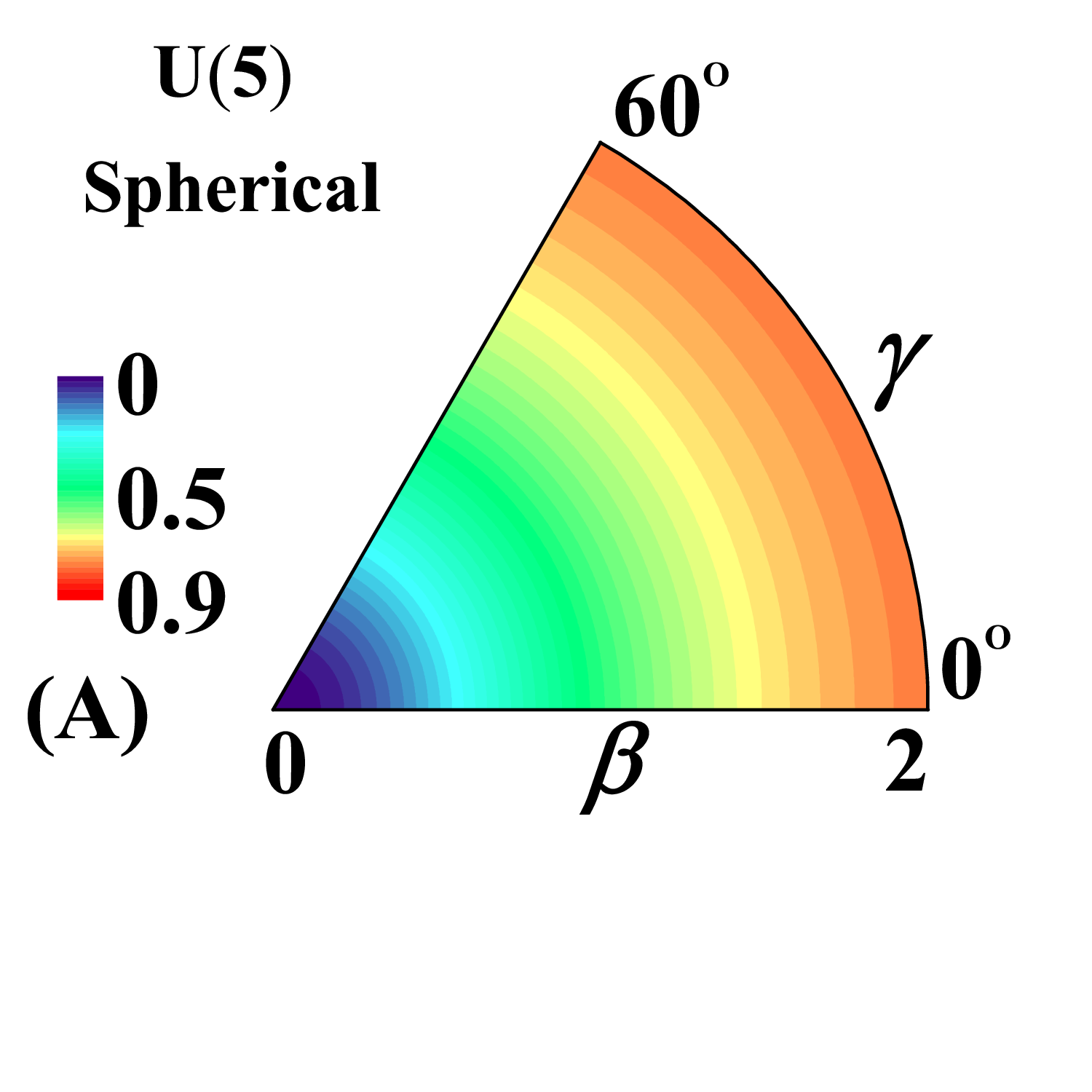}
\includegraphics[scale=0.15]{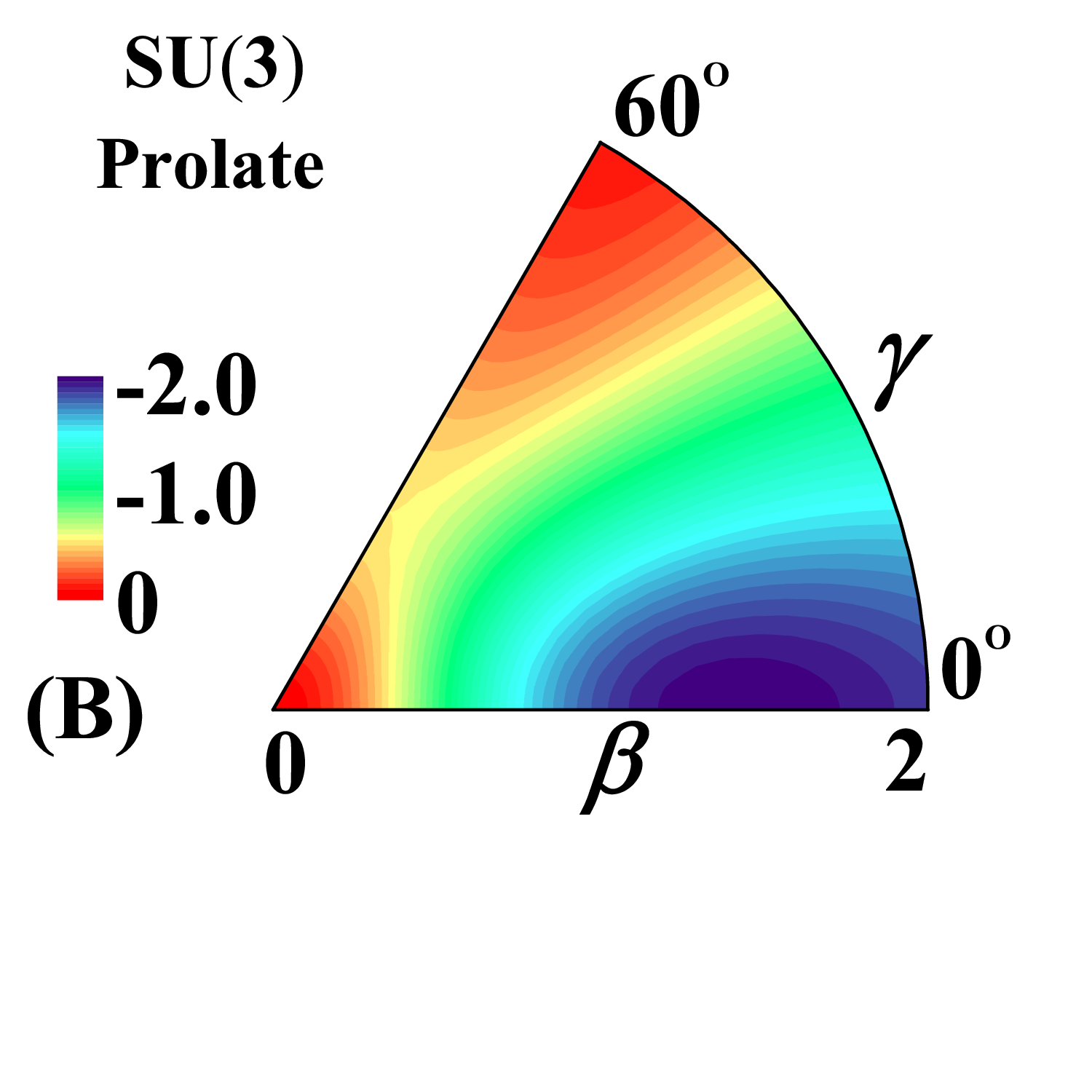}
\includegraphics[scale=0.15]{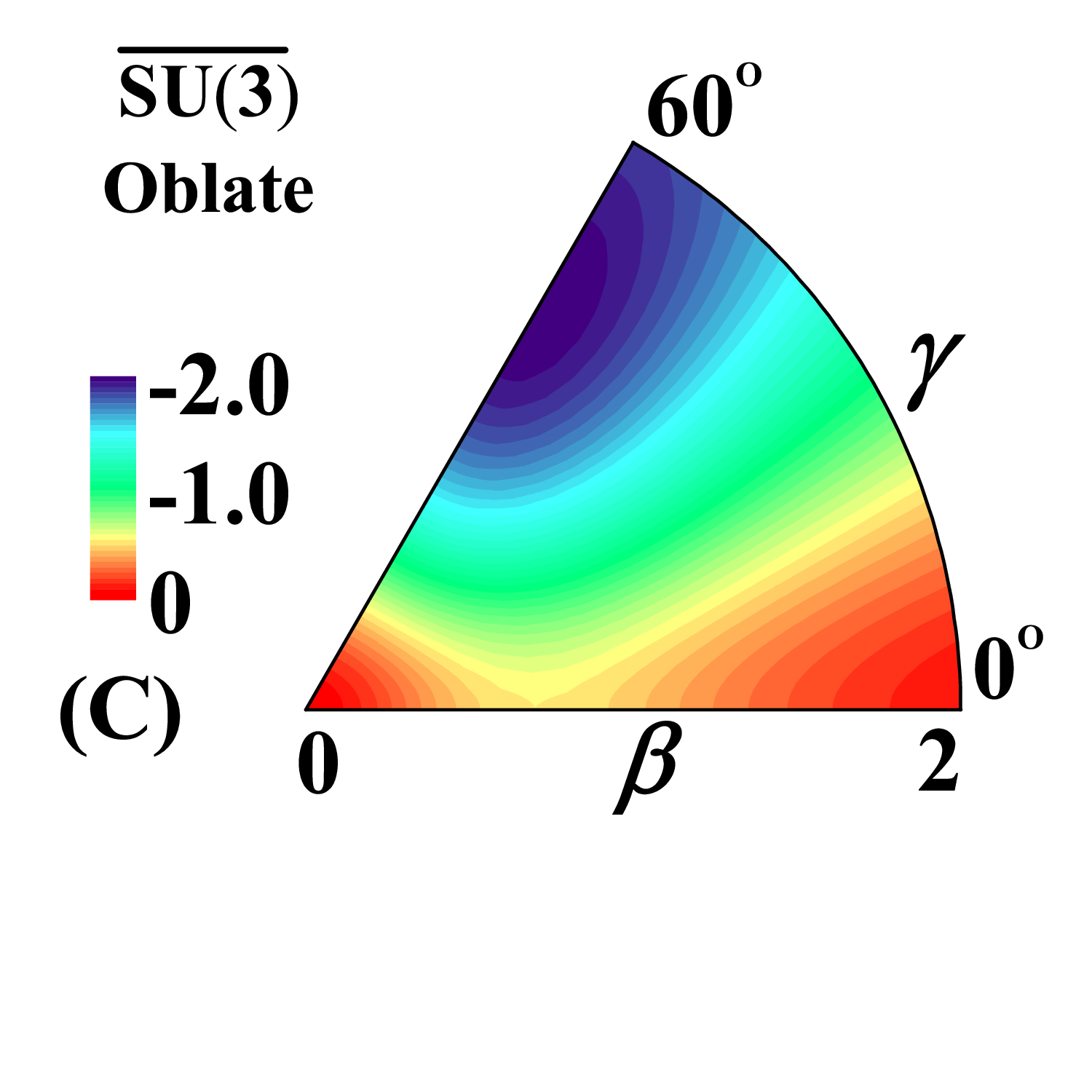}
\includegraphics[scale=0.15]{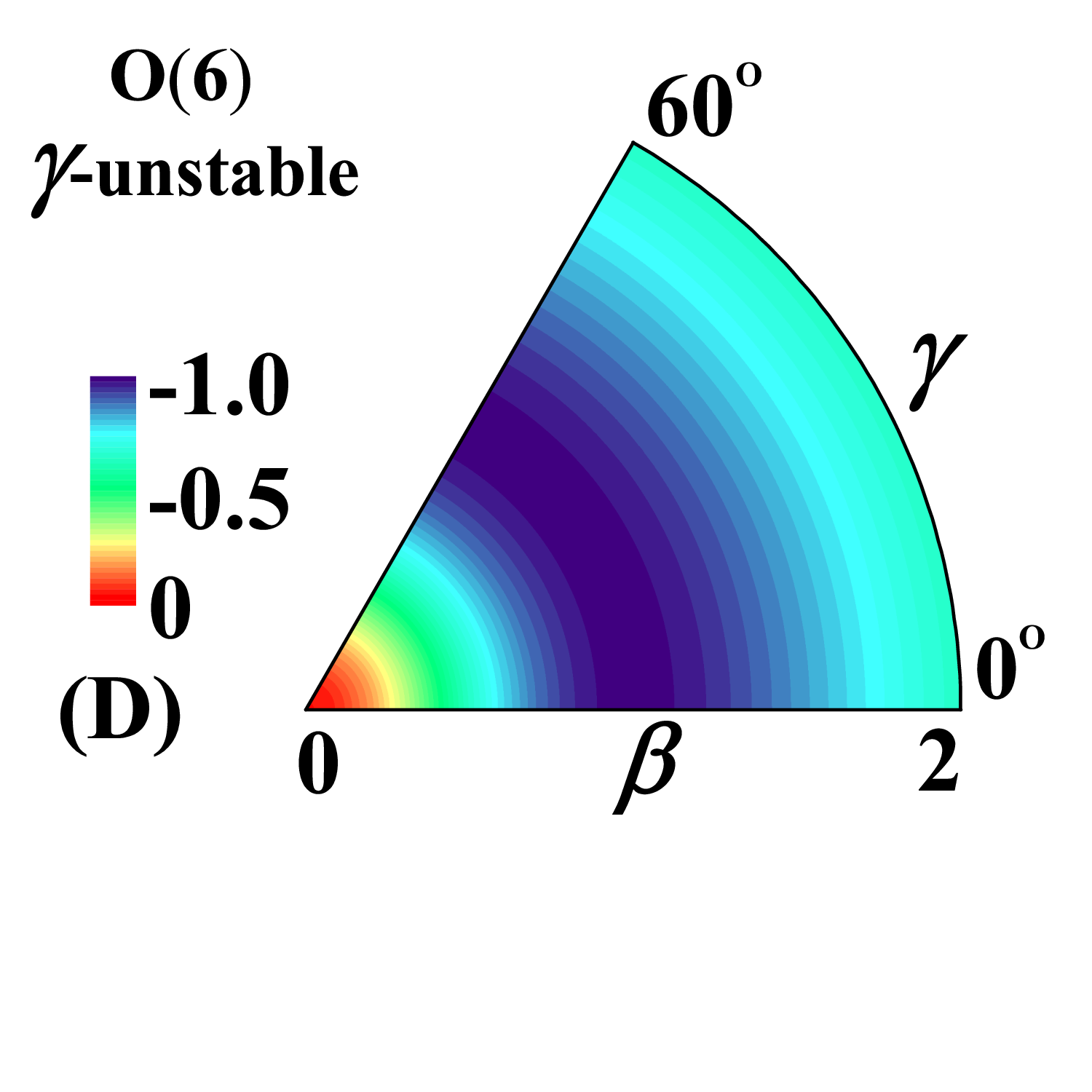}
\caption{Potentials surfaces extracted from Eq.~(\ref{V}) with nonzero parameters
taken as $\varepsilon=1.0$ for U(5), $\kappa=-1.0$
and $\chi=0$ for O(6),
and $\kappa=-1.0$ together with $\chi=\mp\sqrt{7}/2$ for SU(3) (prolate \& oblate shapes).}\label{F1}
\end{center}
\end{figure}

The IBM Hamiltonian is constructed from two kinds of
 boson operators, namely the $s$-boson with $l^\pi = 0^+$ and the
 $d$-boson with $l ^ \pi = 2 ^ +$~\cite{IachelloBook87}.
The total boson number $N$ is
 taken as the number of  valence particle or hole pairs
 for a given
 nucleus. Three dynamical symmetry limits in the IBM are characterized
 by three different group chains of the U(6) group~\cite{IachelloBook87}:

 \begin{eqnarray}
 &&\mathrm{U(6)} \supset \mathrm{U(5)} \supset \mathrm{O(5)} \supset \mathrm{O(3)}\, ,\\
 &&\label{SU3} \mathrm{U(6)} \supset \mathrm{SU(3)} \supset \mathrm{O(3)}\, ,\\
 &&\mathrm{U(6)} \supset \mathrm{O(6)} \supset \mathrm{O(5)} \supset
 \mathrm{O(3)}\, .
 \end{eqnarray}
 Dynamical symmetry associated with each group chain
 corresponds to a typical nuclear shape or collective mode, including
 the spherical vibration in the U(5) limit, the axially-symmetric rotation in the SU(3) limit, and the $\gamma$-unstable
 motion~\cite{Wilets1956} in the O(6) limit. To describe these modes, it is convenient to adopt the consistent-$Q$ Hamiltonian, which is written as~\cite{Warner1983}
\begin{eqnarray}\label{CQ}
\hat{H}_{\mathrm{CQ}}=&\varepsilon\,\hat{n}_d+{\kappa\over{N}}\hat{Q}^\chi\cdot\hat{Q}^\chi
\,
\end{eqnarray}
with
\begin{eqnarray} \label{LQ}
&&\hat{n}_d=d^\dag\cdot\tilde{d},\\ \label{Qx}
&&\hat{Q}_u^\chi =
(d^{\dag} s + s^{\dag} \tilde{d})_u^{(2)} + \chi
(d^{\dag}\times\tilde{d})_u^{(2)}\, ,
\end{eqnarray}
where $\varepsilon,~\kappa,~\chi$ are real parameters, and $N$ is the total  number of the bosons.
The three dynamical symmetry
limits of the consistent-$Q$ Hamiltonian (\ref{CQ})
can be characterized as: the U(5) limit for $\varepsilon>0$ and $\kappa=0$; the O(6) limit for
$\varepsilon=0$, $\kappa<0$ and $\chi=0$; and the SU(3) limit for
$\varepsilon=0$, $\kappa<0$ and $\chi=\pm\sqrt{7}/{2}$.
To elucidate the geometric aspects of the IBM, one can examine its classical limit by employing the coherent state defined by~\cite{IachelloBook87}
\begin{eqnarray}
|\beta, \gamma, N\rangle=N_A[s^\dag + \beta \mathrm{cos} \gamma~
d_0^\dag\ + \frac{1}{\sqrt{2}} \beta \mathrm{sin} \gamma (d_2^\dag +
d_{ - 2}^\dag)]^N |0\rangle\,
\end{eqnarray}
with $N_A=1/\sqrt{N!(1+\beta^2)^N}$. The classical potential
corresponding to $\hat{H}_{\mathrm{CQ}}$ is
thus given by
\begin{eqnarray}\label{V}\nonumber
V_\mathrm{CQ}(\beta, \gamma)&=&\frac{1}{N}\langle\beta, \gamma,
N|\hat{H}_{\mathrm{CQ}}|\beta, \gamma,N\rangle|_{N\rightarrow\infty}\\
\nonumber
&=&\varepsilon\frac{\beta^2}{1+\beta^2}+\kappa\frac{1}{(1+\beta^2)^2}\\
&\times&[4\beta^2- 4\sqrt{\frac{2}{7}}\chi\beta^3 \mathrm{cos3}
\gamma + \frac{2}{7}\chi^2\beta^4]\, .
\end{eqnarray}
To extract the mean-field-type deformation, one can minimize
the potential function by varying $\beta$ and $\gamma$
for the given parameters $\varepsilon,~\kappa$ and $\chi$. The resulting optimal values are denoted by
$\beta_\mathrm{e}$ and $\gamma_\mathrm{e}$, with which one gets
the ground state energy per boson, $E_g=V(\varepsilon,\kappa,\chi,\beta_\mathrm{e},\gamma_\mathrm{e})$.
Using this procedure, one can not only
extract deformations (shapes) of the IBM systems but also identify the associated shape phase transitions~\cite{IachelloBook87}.
To illustrate the dynamical symmetry limits in the IBM, the corresponding potential surfaces are
provided in Fig~\ref{F1}.
It is clearly shown from Fig.~\ref{F1}
that the U(5) potential has its minimum at $\beta_\mathrm{e}=0$ indicating the spherical shape, while the SU(3) ones have
their minima at either $\gamma_\mathrm{e}=0^\circ$ or $\gamma_\mathrm{e}=60^\circ$, which correspond to the prolate (SU(3)) and oblate ($\mathrm{\overline{SU(3)}}$), respectively.
Clearly, triaxial minimum with $0^\circ<\gamma_\mathrm{min}<60^\circ$ cannot be generated by these potentials even for the O(6) one with nonzero $\beta$
as shown in Fig~\ref{F1} (D), in which the potential, instead, manifests a $\gamma$-unstable picture with the potential minimum being independent of the $\gamma$ variable.
In short, the triaxial rotor mode
widely used to interpret nuclear excitations~\cite{Davydov1958}
cannot be produced from
any IBM Hamiltonian up to two-body
interactions~\cite{VC1981}.

\begin{figure}
\begin{center}
\includegraphics[scale=0.3]{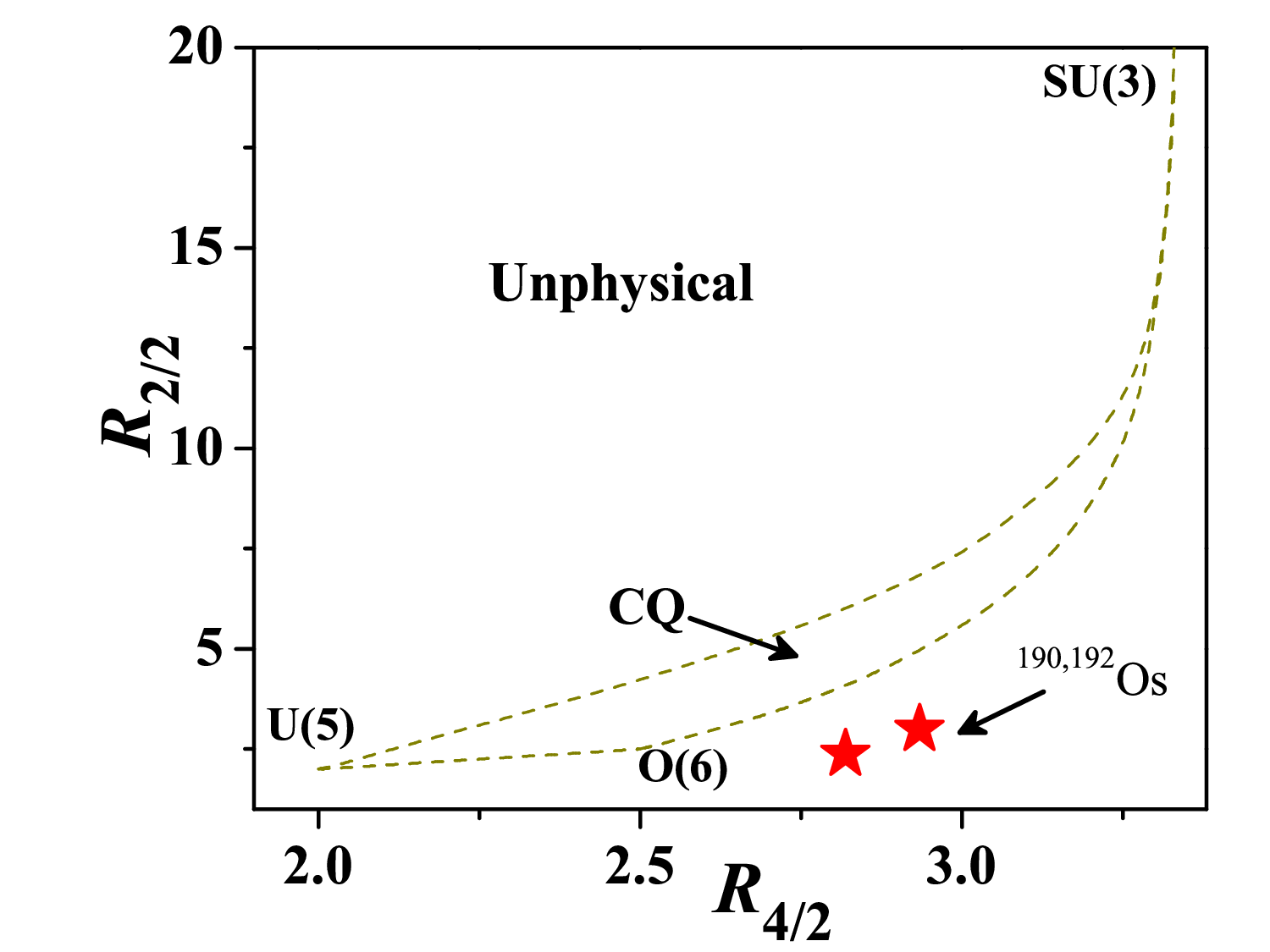}
\caption{Trajectories of $R_{2/2}$ vs $R_{4/2}$
obtained from the consistent-$Q$ (CQ) Hamiltonian with $N=9$.
The red stars represent the experimental values of the ratios for $^{190,192}$Os~\cite{Singh2003,Baglin2012}. No nucleus
is observed to exist in the upper left "unphysical" region above the CQ region surrounded by the dashed curve.}\label{F2}
\end{center}
\end{figure}

In addition to variations in potential surfaces, different modes may present distinct spectral patterns, which can be alternatively identified through specific characteristic quantities.
The simplest ones may be the energy ratios, $R_{4/2}\equiv E(4_1^+)/E(2_1^+)$ and $R_{2/2}\equiv E(2_2^+)/E(2_1^+)$, in which $E(2_2^+)$ is
the band head energy of the $\gamma$- or
quasi-$\gamma$-band
sensitive to the $\gamma$ deformation.
Typically,  $R_{4/2}\approx R_{2/2}\approx2.0$ for the U(5) mode; $R_{4/2}\approx R_{2/2}\approx2.5$ for the O(6) mode; and  $R_{4/2}\approx3.3$ but with $R_{2/2}\gg R_{4/2}$  for the SU(3) or $\mathrm{\overline{SU(3)}}$ mode~\cite{Jolie2001}.
The trajectories of $R_{4/2}$ vs $R_{2/2}$ calculated from $\hat{H}_{\mathrm{CQ}}$ with $N=9$ are provided in FIG.~\ref{F2}. In the calculation, the parameters have been expressed
as $\varepsilon=(1-\eta)$ and $\kappa=-2\eta$ for convenience, with which the evolution among different modes in the IBM is described by
the control parameter
$\eta\in[0,1]$ and $\chi\in[0,-\sqrt{7}/2]$~\cite{Iachello2004}.
Here, we only discuss the cases with $\chi\leq0$ as the Hamiltonian possesses the $Z_2$ symmetry with $\chi\leftrightarrow-\chi$~\cite{Iachello2004}.
It is shown in FIG.~\ref{F2}
that there is a monotonic change in $R_{2/2}$ as a function of $R_{4/2}$ in between any two modes. Such a monotonic evolution of $R_{2/2}$ against $R_{4/2}$ may be kept for any given $\chi$.
Particularly, the narrow region
surrounded by the dashed curve
is produced from the consistent-$Q$ Hamiltonian.
Moreover, it is shown that the area below the CQ-region
with small $R_{2/2}$ but large $R_{4/2}$ as typically
marked  by the experimental values of $^{190,192}$Os
is beyond the scope of the consistent-$Q$ Hamiltonian. Interestingly, this scenario aligns well with the characteristics of triaxial rotor modes, providing strong impetus for the development of the IBM representation of such modes. This endeavor promises to enhance our understanding of the intricate interplay among different collective modes within a unified framework.

\begin{center}
\vskip.2cm\textbf{3. The IBM realization of triaxial rotor modes}
\end{center}\vskip.2cm

The algebraic approach
established in the shell model description of a quantum rotor
~\cite{Leschber1987,Castanos1988} was employed
to construct
the triaxial rotor mode in the IBM ~\cite{Smirnov2000} with the Hamiltonian constructed
from the symmetry-conserving operators
of the $\mathrm{SU(3)}\supset\mathrm{SO(3)}$ integrity
basis~\cite{Vanden1985}. From a group (algebra) theory point of
view~\cite{Ui1970}, the su(3) algebraic relation
in the large-$N$
limit will contract to the Lie algebra
$\mathrm{t_5}\oplus\mathrm{so(3)}$ of a quantum rotor. Based on the early works \cite{Leschber1987,Castanos1988,Smirnov2000},
an exact mapping between the triaxial rotor and its IBM image was then established in the SU(3) limit of the IBM~\cite{Zhang2014}.

The Hamiltonian of a rigid quantum rotor is
given by~\cite{Davydov1958,Wood2004}
\begin{equation}\label{Hr}
\hat{H}_{\mathrm{rot}}=A_1\hat{L}_1^2+A_2\hat{L}_2^2+A_3\hat{L}_3^2\, ,
\end{equation}
where $L_\alpha$ is the projection of the angular momentum onto the
$\alpha$-th body-fixed principal axis and $A_\alpha$ ($\alpha=1,2,3$) is the
corresponding inertia parameter.
On the other hand, one
can construct three frame-independent scalars~\cite{Leschber1987,Castanos1988},
\begin{eqnarray}\label{LXX1}
&&\hat{L}^2=\hat{L}_1^2+\hat{L}_2^2+\hat{L}_3^2\, ,\\ \label{LXX2}
&&\hat{X}_3^c=\sum_{\alpha\beta}\hat{L}_\alpha \hat{Q}_{\alpha\beta}^c \hat{L}_\beta=\lambda_1\hat{L}_1^2+\lambda_2\hat{L}_2^2+\lambda_3\hat{L}_3^2\, ,\\ \label{LXX3}
&&\hat{X}_4^c=\sum_{\alpha\beta\gamma}\hat{L}_\alpha \hat{Q}_{\alpha\beta}^c\hat{Q}_{\beta\gamma}^c \hat{L}_\gamma=\lambda_1^2\hat{L}_1^2+\lambda_2^2\hat{L}_2^2+\lambda_3^2\hat{L}_3^2\, ,
\end{eqnarray}
where the expressions shown on the right-hand-side
of  (\ref{LXX1}){--}(\ref{LXX3})
are obtained in the body-fixed
principal axes system.
The rotor model operator $\hat{L}_\alpha$ and $\hat{Q}_{\alpha\beta}^c$ are the cartesian form of
$\hat{L}_u$ and $\hat{Q}_u^c$ defined as
\begin{eqnarray}\label{L}
&&\hat{L}_u=\int\rho(\vec{r})(\vec{r}\times\vec{v})_u~d\tau\, ,\\ \label{Qc}
&&\hat{Q}_u^c=\sqrt{16\pi/5}\int\rho(\vec{r})r^2Y_{2u}(\Omega)~d\tau\, ,
\end{eqnarray}
where $\rho(\vec{r})$ represents nuclear mass density and the integration is
over the whole nuclear volume. With the spherical tensor formulas,
\begin{eqnarray}
&&\hat{L}_0=\hat{L}_z,\\
&&\hat{L}_{\pm1}=\mp\frac{1}{\sqrt{2}}(\hat{L}_x\pm i\hat{L}_y),\\
&&\hat{Q}_0^c=3\hat{Q}_{zz}^c,\\
&&\hat{Q}_{\pm1}^c=\mp\sqrt{6}(\hat{Q}_{xz}^c\pm i\hat{Q}_{yz}^c),\\
&&\hat{Q}_{\pm2}^c=\sqrt{\frac{3}{2}}(\hat{Q}_{xx}^c-\hat{Q}_{yy}^c\pm 2i\hat{Q}_{xy}^c)\, ,
\end{eqnarray}
the scalars given in (\ref{LXX1})-(\ref{LXX3}) can be reexpressed as
\begin{eqnarray}\label{a}
&&\hat{L}^2=\sqrt{5}(\hat{L}\times \hat{L})^{(0)}\, ,\\ \label{b}
&&\hat{X}_3^c=\frac{\sqrt{30}}{6}(\hat{L}\times \hat{Q}^c\times \hat{L})^{(0)}\, ,\\ \label{c}
&&\hat{X}_4^c=\frac{5}{18}(\hat{L}\times \hat{Q}^c)^{(1)}\cdot(\hat{L}\times \hat{Q}^c)^{(1)}\, .
\end{eqnarray}
Based on Eq.~(\ref{LXX1})-(\ref{LXX3}), one can derive that
\begin{equation}\label{Ia}
\hat{L}_{\alpha}^2=[(\lambda_1\lambda_2\lambda_3)\hat{L}^2+\lambda_\alpha^2\hat{X}_3^c+\lambda_\alpha
\hat{X}_4^c]/(2\lambda_\alpha^3+\lambda_1\lambda_2\lambda_3)\, ,
\end{equation}
where $\lambda_\alpha$ with $\alpha=1,2,3$ (or $x,y,z$) are the expectation values of the quadrupole matrix in the principal-axes system
with $\langle \hat{Q}_{\alpha\beta}^c\rangle=\lambda_\alpha\delta_{\alpha\beta}$ satisfying $\lambda_1+\lambda_2+\lambda_3=0$.
Thus, the rotor Hamiltonian (\ref{Hr}) can be exactly rewritten as
\begin{equation}\label{HLQ}
\hat{H}_{\mathrm{rot}}=a\hat{L}^2+b\hat{X}_3^c+c\hat{X}_4^c\,
\end{equation}
with
\begin{eqnarray}\label{abc}\nonumber
&&a=\sum_\alpha a_\alpha\,
A_\alpha,~~~~a_\alpha=\lambda_1\lambda_2\lambda_3/D_\alpha\, ,\\
&&b=\sum_\alpha b_\alpha\,
A_\alpha,~~~~b_\alpha=\lambda_\alpha^2/D_\alpha\, ,\\\nonumber
&&c=\sum_\alpha c_\alpha\,
A_\alpha,~~~~c_\alpha=\lambda_\alpha/D_\alpha\, ,
\end{eqnarray}
and
\begin{equation}
D_\alpha=2\lambda_\alpha^3+\lambda_1\lambda_2\lambda_3\, .
\end{equation}
Since the rotor Hamiltonian (\ref{HLQ}) is
frame independent,
an SU(3) algebraic realization of the triaxial rotor
can be achieved
by replacing $\hat{L}$ and $\hat{Q^c}$
in the Hamiltonian with the SU(3)
group generators, $\hat{L}$ and $\hat{Q}$~\cite{Leschber1987,Castanos1988}.
Accordingly, the values of $\lambda_\alpha$ can
be evaluated  from the linear relations
suggested in \cite{Castanos1988} with
\begin{eqnarray}\label{lmmu}\nonumber
 &&\lambda_1=-(\lambda-\mu)/3\, ,\\
 &&\lambda_2=-(\lambda+2\mu+3)/3\, ,\\ \nonumber
 &&\lambda_3=(2\lambda+\mu+3)/3\, .
\end{eqnarray}
Here, the quantum numbers $(\lambda,\mu)$ are the labels of the irreducible representations (IRREPS)
of the SU(3) group, with which
 the deformation parameter $\gamma$
 can be expressed as~\cite{Castanos1988}
\begin{equation}\label{gamma}
\gamma_\mathrm{S}=\mathrm{tan}^{-1}\Big(\frac{\sqrt{3}(\mu+1)}{2\lambda+\mu+3}\Big)\, .
\end{equation}

With the mapping shown above,
the rotor image can
be established in any nuclear model with the SU(3) symmetry.
However, the Hamiltonian (\ref{HLQ})
does not contain any information of the nuclear ground-state
deformation, which will be produced by adding a static (intrinsic) part to the Hamiltonian~\cite{Smirnov2000}.
As a result, the full IBM Hamiltonian for triaxial rotor
can be constructed with
two parts~\cite{Zhang2014,Zhang2022},
\begin{eqnarray}\label{Tri}
\hat{H}_{\mathrm{Tri}}=\hat{H}_\mathrm{S}+\hat{H}_\mathrm{D}\,
\end{eqnarray}
with
\begin{eqnarray}\label{HS}
&\hat{H}_\mathrm{S}=\frac{a_1}{N}\hat{C}_2[\mathrm{SU(3)}]+\frac{a_2}{N^3}\hat{C}_2[\mathrm{SU(3)}]^2+\frac{a_3}{N^2}\hat{C}_3[\mathrm{SU(3)}],\\
\label{HD}
&\hat{H}_\mathrm{D}=
t_1\hat{L}^2+t_2(\hat{L}\times\hat{Q}\times\hat{L})^{(0)}+t_3(\hat{L}\times\hat{Q})^{(1)}\cdot(\hat{L}\times\hat{Q})^{(1)}\,~~
\end{eqnarray}
where $a_i$ and $t_i$ $(i=1,2,3)$ are real parameters,
$\hat{L}$ and $\hat{Q}$ denote the angular momentum and
quadrupole momentum operators defined in the SU(3) basis
with
\begin{eqnarray}\label{SQIL}
&&\hat{L}_u=\sqrt{10}(d^\dag\times\tilde{d})_u^{(1)}\, ,\\ \label{SQI}
&&\hat{Q}_u=2\sqrt{2}[(d^\dag\times\tilde{s}+s^\dag\times\tilde{d})_u^{(2)}-\frac{\sqrt{7}}{2}(d^\dag\times\tilde{d})_u^{(2)}]\, .
\end{eqnarray}
Clearly, the dynamical part $\hat{H}_\mathrm{D}$
is just taken from the rotor image $\hat{H}_\mathrm{rot}$ given in (\ref{HLQ}) with the parameters
\begin{equation}
t_1=a,~~t_2=\frac{\sqrt{30}}{6}b,~~t_3=\frac{5}{18}c\, ,
\end{equation}
while the
static (intrinsic) part $\hat{H}_\mathrm{S}$ is employed to determine the ground-state deformation, which involves
the SU(3) Casimir operators
defined by
\begin{eqnarray}\label{C2}
&\hat{C}_2[\mathrm{SU(3)}]=\frac{1}{4}\hat{Q}\cdot\hat{Q}+\frac{3}{4}\hat{L}^2,\\
\label{C3}
&\hat{C}_3[\mathrm{SU(3)}]=-\frac{\sqrt{70}}{72}(\hat{Q}\times\hat{Q}\times\hat{Q})^{(0)}
-\frac{\sqrt{30}}{8}(\hat{L}\times\hat{Q}\times\hat{L})^{(0)}.~~~~~~
\end{eqnarray}
Eigenvalues of the Casimir operators can be expressed in terms of the SU(3) IRREP with
\begin{eqnarray}
&\langle\hat{C}_2[\mathrm{SU(3)}]\rangle=\lambda^2+\mu^2+3\lambda+3\mu+\lambda\mu,\\
&\langle\hat{C}_3[\mathrm{SU(3)}]\rangle=\frac{1}{9}(\lambda-\mu)(2\lambda+\mu+3)(\lambda+2\mu+3)\,
.
\end{eqnarray}
The ground-state energy of the triaxial Hamiltonian (\ref{Tri}) is then
obtained by $E_g=\langle\hat{H}_\mathrm{S}\rangle_g=f(\lambda,\mu)$ evaluated at the optimal values
$(\lambda_0,\mu_0)$, which is in turn determined by the parameters $a_i$, as $\hat{H}_\mathrm{D}$ contributes nothing to the ground-state energy with $L=0$.
Accordingly, the static triaxiality is determined from $(\lambda_0,\mu_0)$
via Eq.~(\ref{gamma}).
\begin{figure*}
\begin{center}
\includegraphics[scale=0.21]{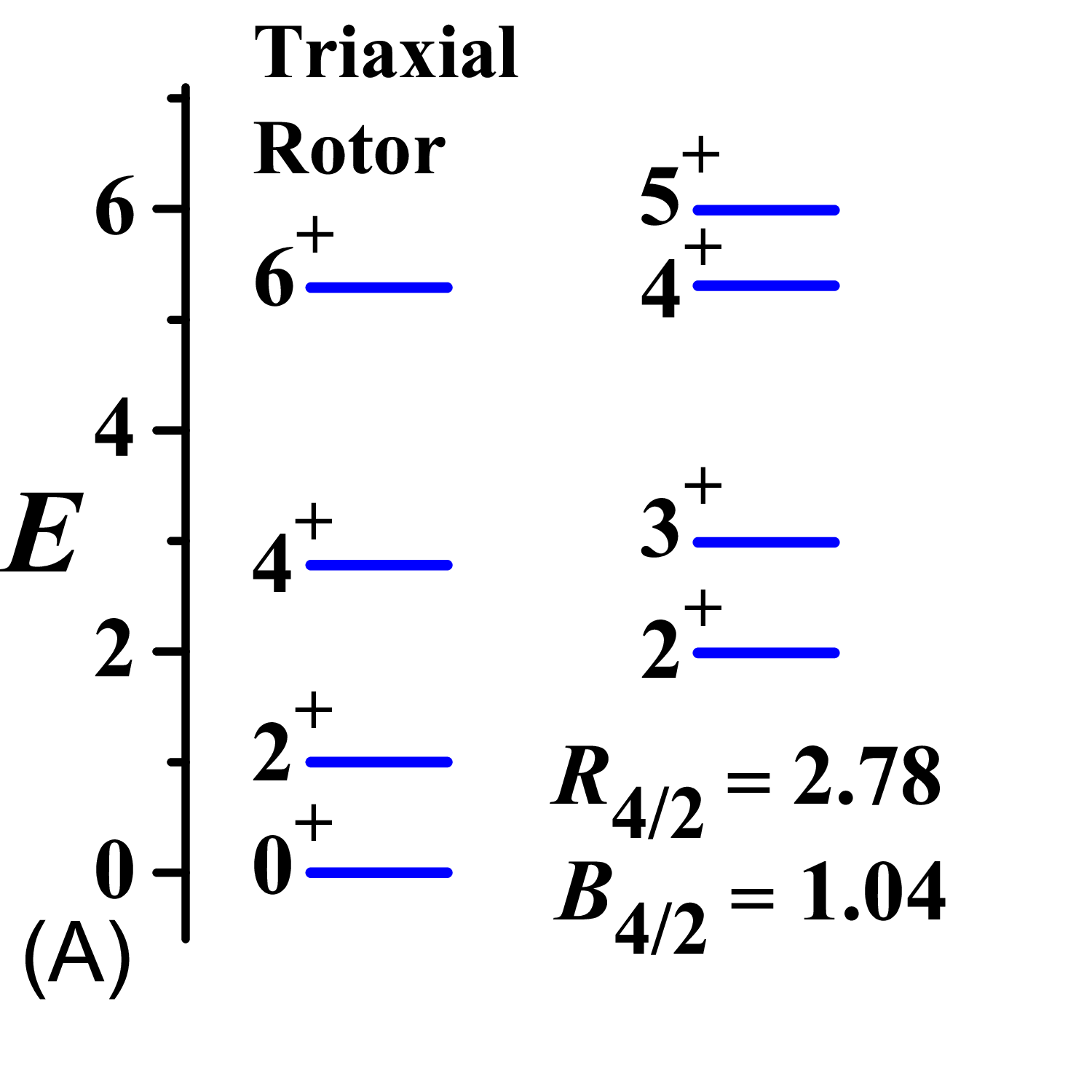}
\includegraphics[scale=0.21]{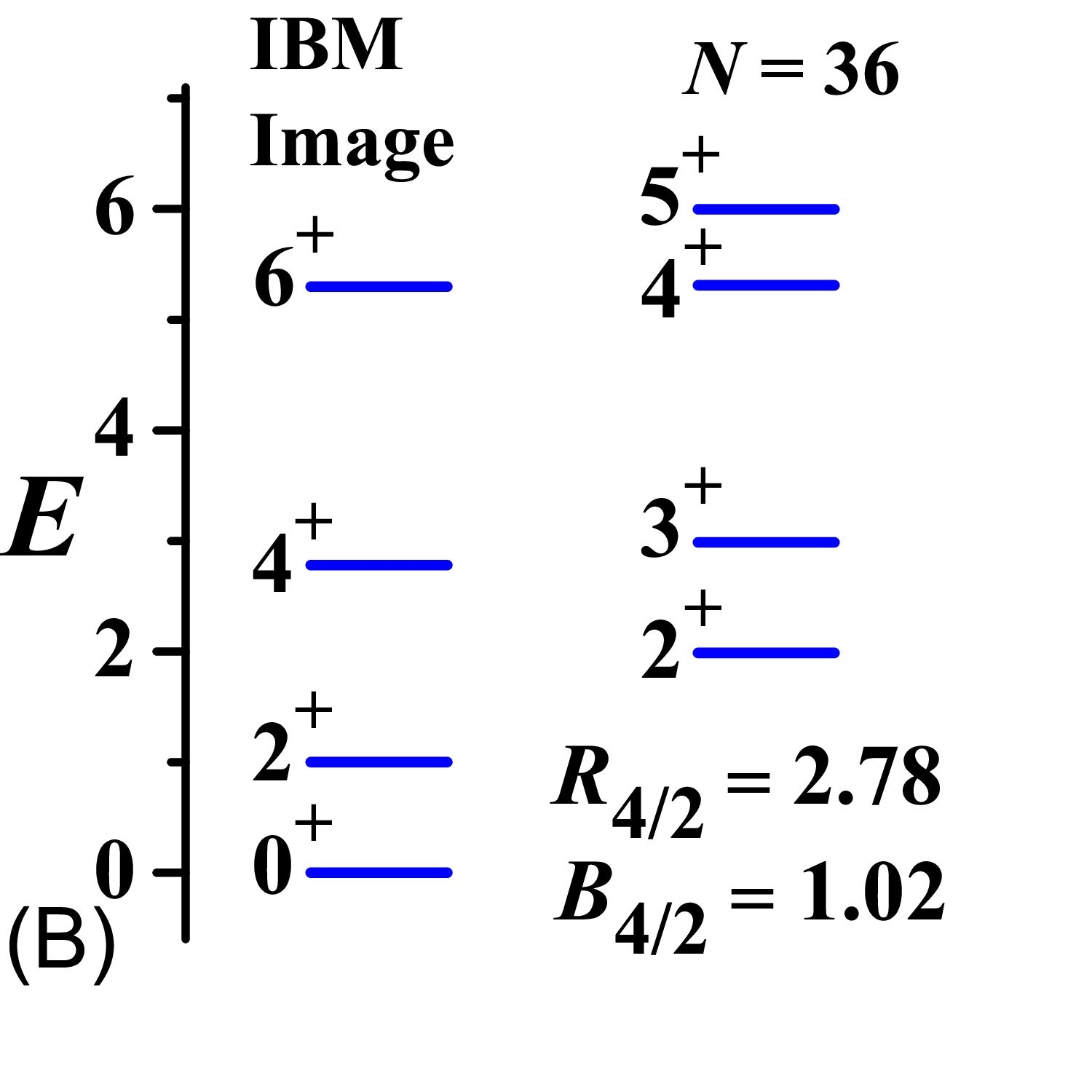}
\includegraphics[scale=0.21]{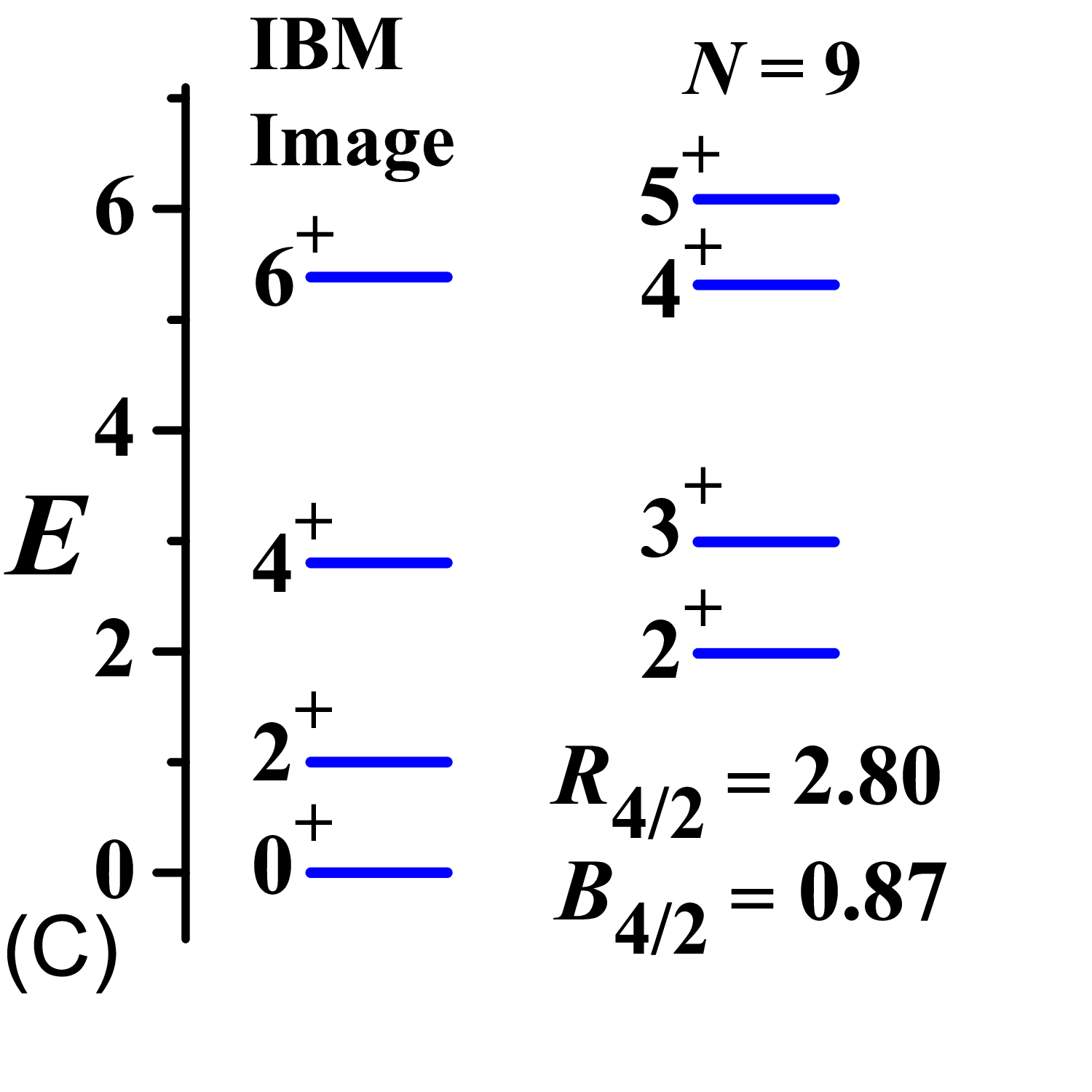}
\caption{(A) The level pattern solved from the rotor Hamiltonian (\ref{Hr}) with $A_1:A_2:A_3=3:1:4$ is shown, where all the levels have been normalized to $E(2_1^+)=1.0$. (B) The corresponding results solved from the IBM image described by (\ref{Tri}) with the boson number $N=36$ and other parameters illustrated in the text. (C) The same as in (B) but for $N=9$. In addition, the typical ratios are defined as $R_{4/2}=\frac{E(4_1^+)}{E(2_1^+)}$ and $B_{4/2}=\frac{B(E2;4_1^+\rightarrow2_1^+)}{B(E2;2_1^+\rightarrow0_1^+)}$.}\label{F3}
\end{center}
\end{figure*}

To examine the validity of the designed IBM realization of the rotor model, the results obtained
directly from
the rotor Hamiltonian (\ref{Hr}) with $A_1:A_2:A_3=3:1:4$, which corresponds to a
very asymmetric situation~\cite{Wood2004}, are provided as
examples to compare with those obtained
from its IBM image described by (\ref{Tri}).
In the IBM calculation,
two cases with
$N=36$ and $N=9$ are considered,
respectively.
In addition,
it is assumed that the asymmetric rotor dynamics
is produced from an $N$-boson system with
$\gamma_{~{\mathrm{S}}}=30^\circ$ deformation,
which requires $(\lambda_0,~\mu_0)=(2N/3,~2N/3)$ according to Eq.~(\ref{gamma}).
To yield the corresponding ground-state IRREP, the IBM parameters are set by choosing $a_1:a_2:a_3=-\frac{27+10N}{3N}:1:1$. The other parameters are then determined by the mapping scheme
described above with
$t_1=3.0,~t_2=0.0548,~t_3=-0.00022$ for $N=36$
and  $t_1=3.0,~t_2=0.1956,~t_3=-0.00283$ for $N=9$.
The $E2$ operator in
the IBM image is simply chosen as
the quadrupole operator $\hat{Q}$ defined in (\ref{SQI}),
while that in the rotor model
is given as
\begin{equation}
T_u^{E2}=\sqrt{\frac{5}{16\pi}}Q_0[\mathrm{cos}\gamma \, D_{u,0}+\frac{1}{\sqrt{2}}\mathrm{sin}\gamma\,(D_{u,2}^{(2)}+D_{u,-2}^{(2)})]\, ,
\end{equation}
where $Q_0$ represents the intrinsic charge quadrupole moment.
Form FIG.~\ref{F3},
one can find that the spectral pattern generated by the triaxial rotor model
can be well reproduced by its the IBM image for both $N=36$ and $N=9$. The results confirmed that the rotor modes can indeed
be produced in the IBM framework.
Meanwhile, one can
observe that the $B(E2)$ anomaly
characterized by $B_{4/2}<1.0$
appears in the IBM image at relatively small $N$
as shown in FIG.~\ref{F3}(C),
which means that the finite-$N$ effect
on the triaxial rotor modes
may bring some unconventional feature
that is never seen in
other collective modes.
It will be shown that
such a novel feature can be applied to
explain the anomalous $E2$ transition
phenomena in the neutron-deficient nuclei.

\begin{figure}
\begin{center}
\includegraphics[scale=0.3]{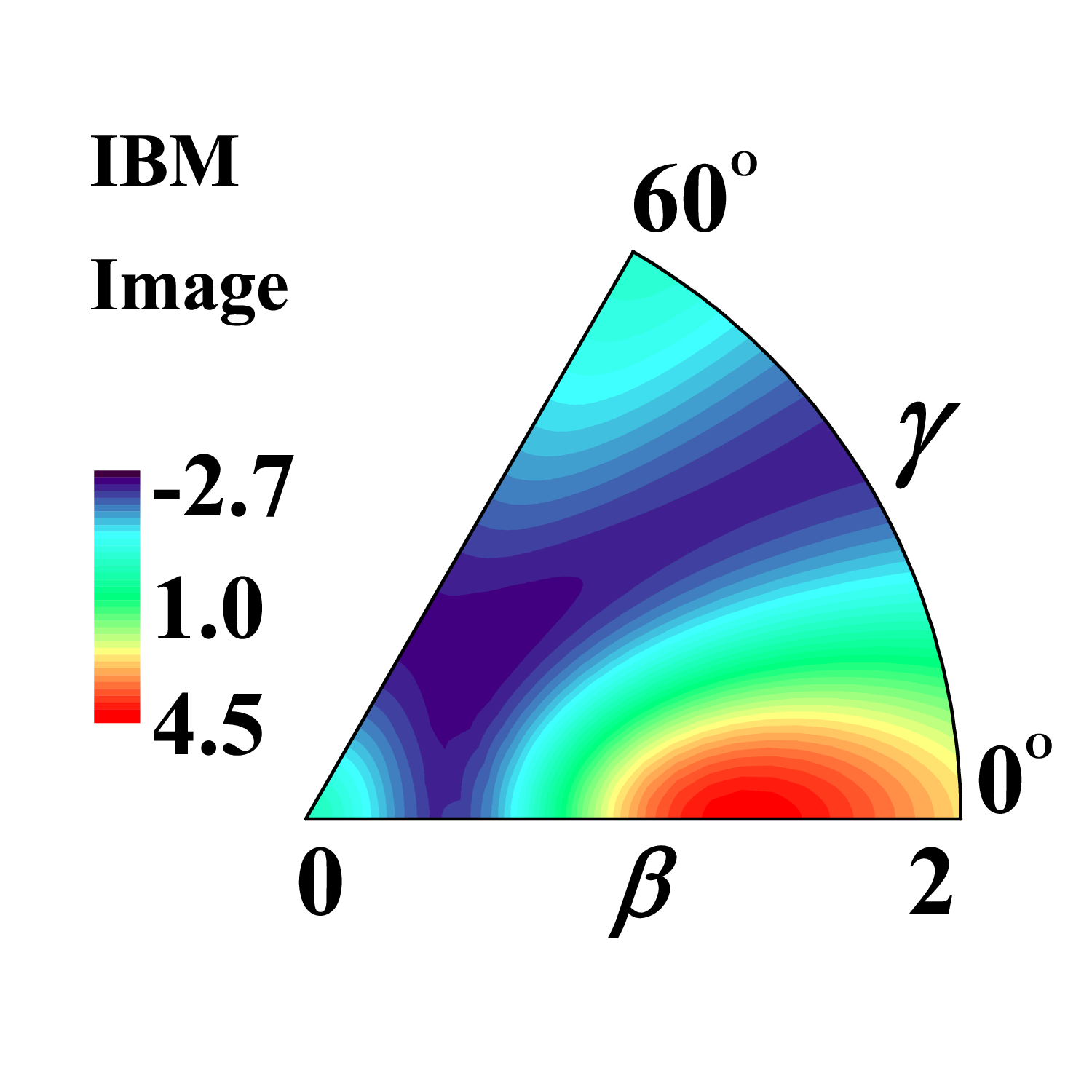}
\caption{The potential surface produced from (\ref{VS}) with $a_1:a_2:a_3=-\frac{27+10N}{3N}:1:1$ corresponding to the IBM image shown in FIG.~\ref{F3} but for $N\rightarrow\infty$.}\label{F4}
\end{center}
\end{figure}

To further understand the IBM image of the rotor model at the mean-filed level, one
may work out
the potential function corresponding to $\hat{H}_\mathrm{S}$ using
the coherent state method~\cite{VC1981}, which
is given by
\begin{eqnarray}\label{VS}\nonumber
V_{\,\mathrm{S}}(\beta,\gamma)&=&\frac{1}{N}\langle\beta, \gamma,
N|\hat{H}_{\mathrm{S}}|\beta, \gamma,N\rangle|_{N\rightarrow\infty}\\
\nonumber
&=&a_1\frac{\beta^2}{(1+\beta^2)^2}\Big[8+4\sqrt{2}\beta\mathrm{cos}(3\gamma)+\beta^2\Big]\\
\nonumber
&+&a_2\frac{\beta^4}{(1+\beta^2)^4}\Big[64+32\beta^2+\beta^4+16\beta^2\mathrm{cos}(6\gamma)\\
\nonumber &+&8\sqrt{2}(8\beta+\beta^3)\mathrm{cos}(3\gamma)\Big]\\
\nonumber
&+&a_3\frac{2\beta^3}{9(1+\beta^2)^3}\Big[24\beta+16\sqrt{2}\mathrm{cos}(3\gamma)\\
 &+&6\sqrt{2}\beta^2\mathrm{cos}(3\gamma)
+\beta^3\mathrm{cos}(6\gamma)\Big]\, .
\end{eqnarray}
The potential
$V_{\,\mathrm{S}}(\beta,\gamma)$
simultaneously describes
the classical limit of both $\hat{H}_{\mathrm{S}}$ and
$\hat{H}_{\mathrm{Tri}}$,
because
the potential for the dynamic part Hamiltonian $\hat{H}_{\mathrm{D}}$
disappears in the large-$N$ limit
through setting an $N$-dependent parameter
form~\cite{Zhang2022}.
With
parameters $a_1:a_2:a_3=-\frac{27+10N}{3N}:1:1$ for the IBM image,
the obtained classical potential surface with $N\rightarrow\infty$ is shown in FIG.~\ref{F4}, which
clearly shows
that the potential surface is completely different from those
presented in FIG.~\ref{F1}.
Specifically, it is shown that
the potential minimum
appears near $\gamma\sim40^\circ$.
This value is close to the one for $\gamma_{~{\mathrm{S}}}$
solved from the formula (\ref{gamma}),
which is more convenient
to quantify
triaxiality for finite-$N$ systems.
Hence, it is clealy shown that
large triaxial deformation can
be produced at the mean-field level under the
given parameters.
In fact, the emergent triaxial deformation can be directly understood from the potential function (\ref{VS}), in which the terms like $\mathrm{cos(6\gamma)}$, which may lead to an asymmetric minimum~\cite{VC1981}, are involved in contrast to that given in (\ref{V}). It is thus confirmed that the IBM image of the rotor model indeed yields the novel modes that are different from those described by the consistent-$Q$ Hamiltonian. Therefore,
the IBM description
can be extended in this way
to the triaxial rotor region corresponding to
the lower right region of FIG.~\ref{F2}.

\begin{center}
\vskip.2cm\textbf{4. Dynamical mixing of different modes}
\end{center}\vskip.2cm

\begin{figure*}
\begin{center}
\includegraphics[scale=0.2]{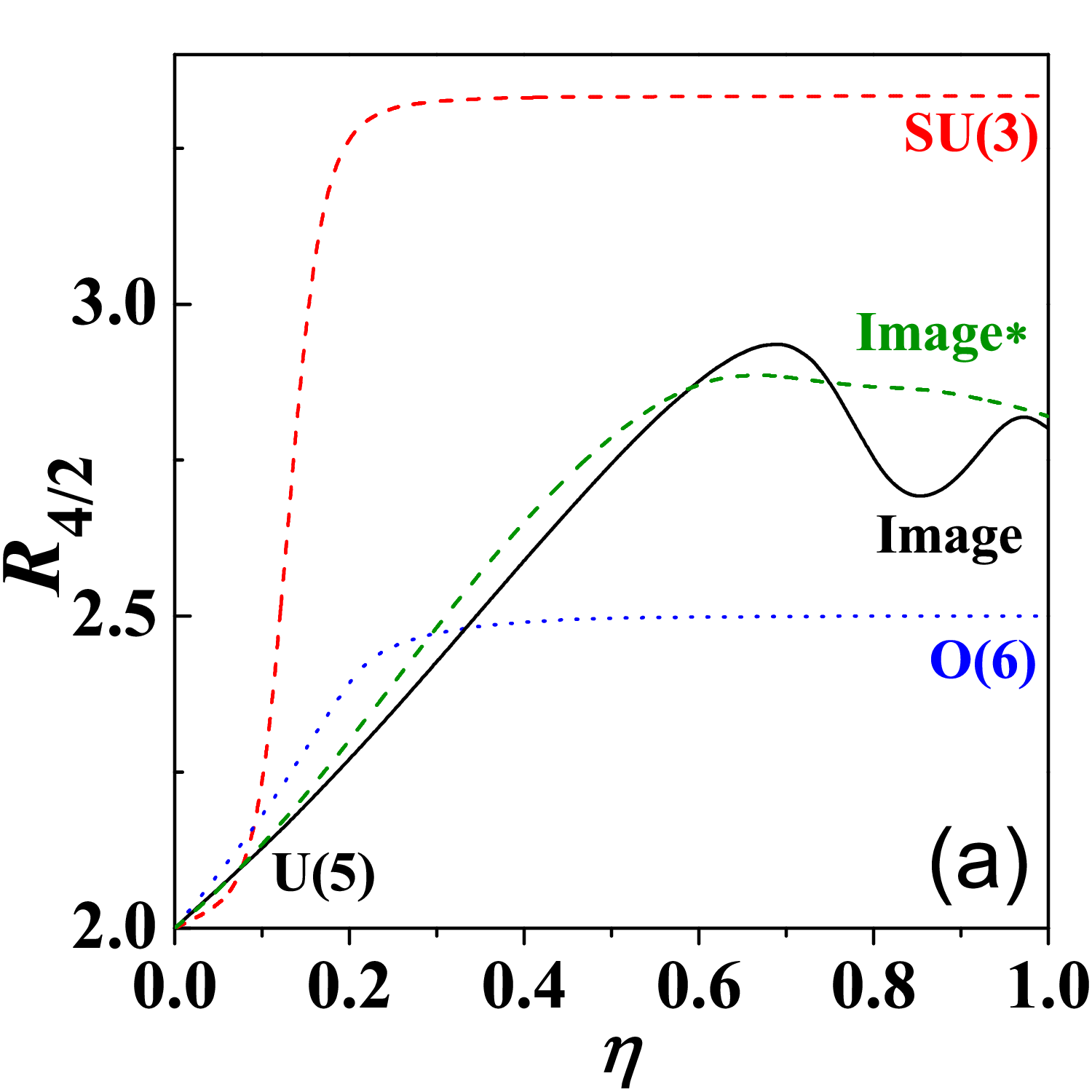}
\includegraphics[scale=0.2]{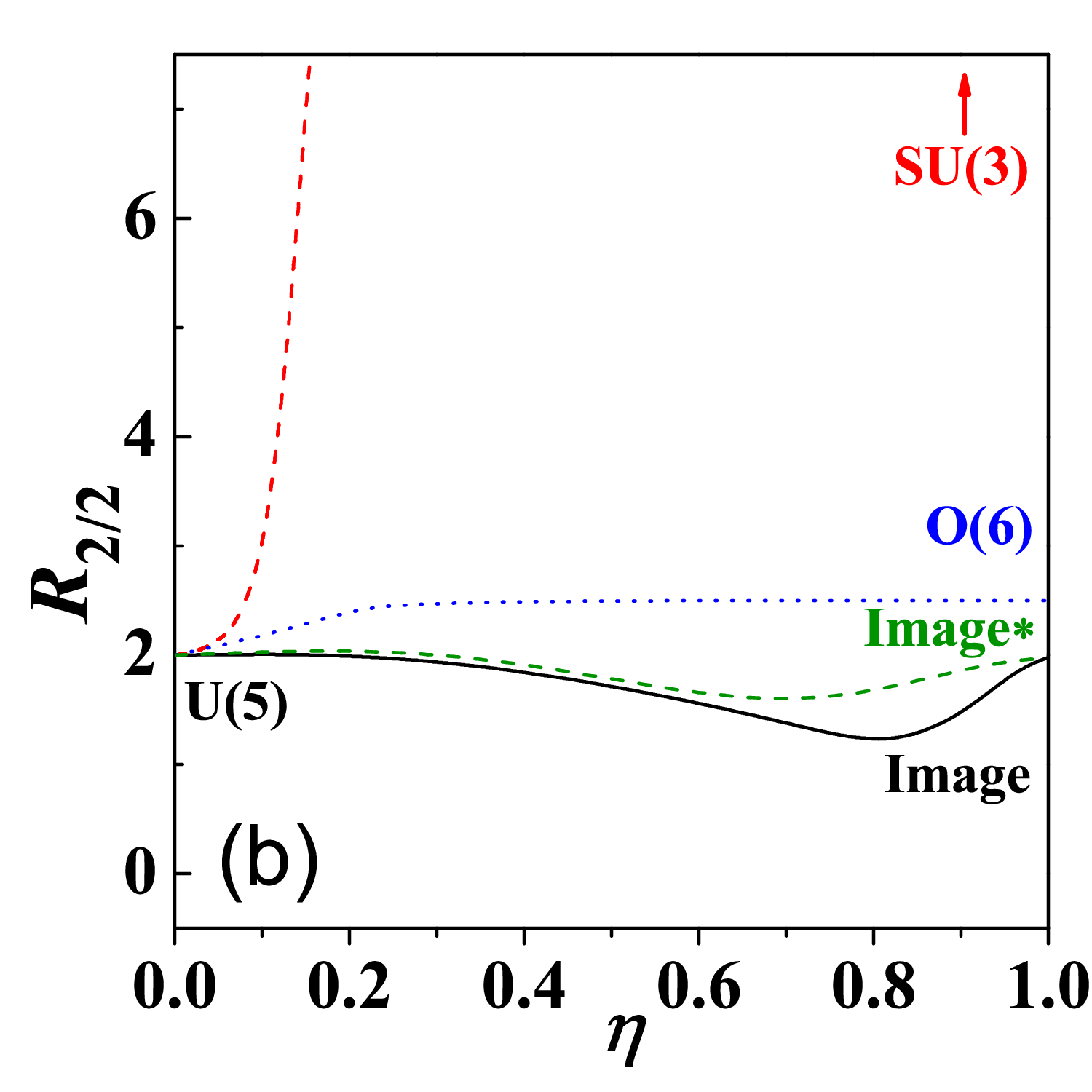}
\includegraphics[scale=0.2]{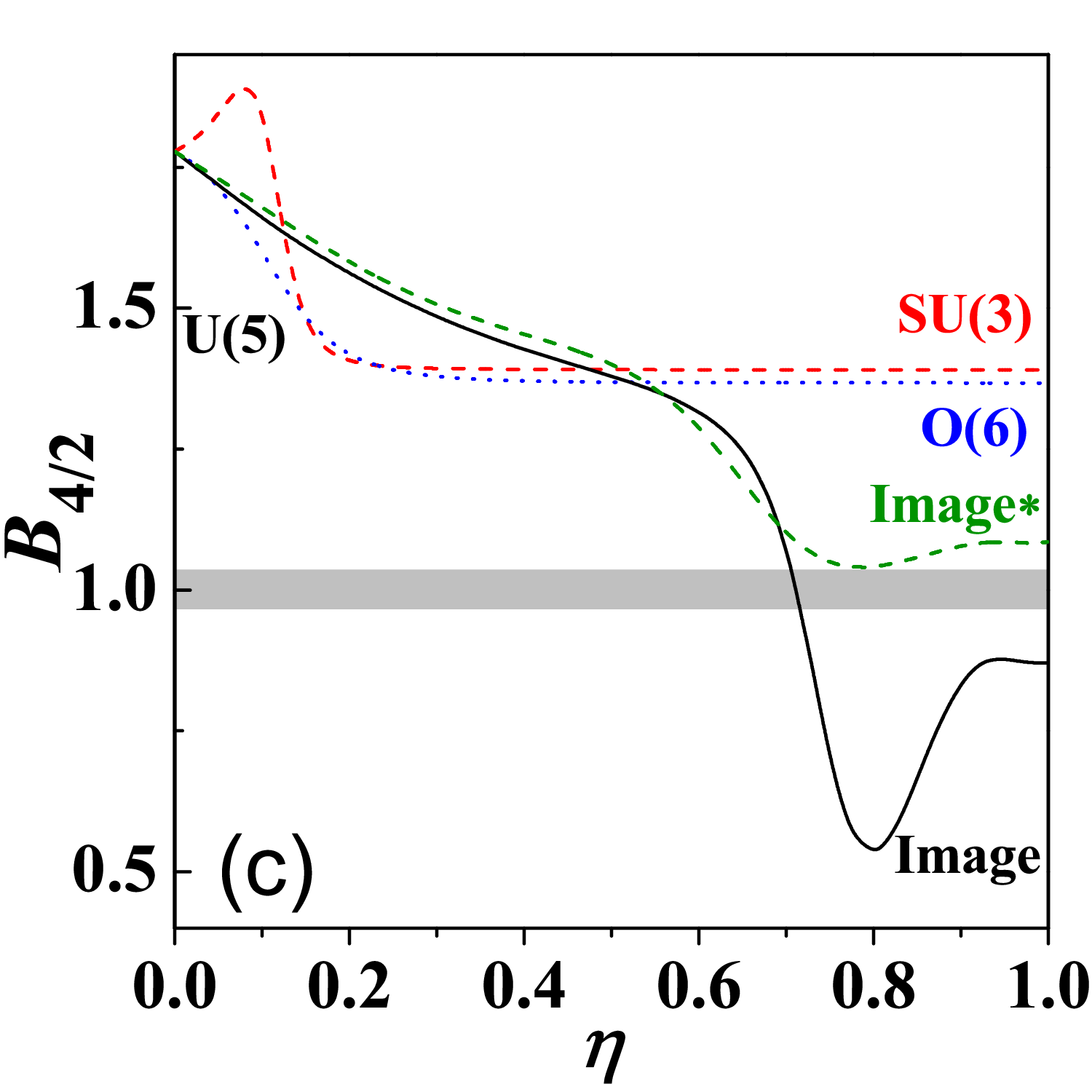}
\includegraphics[scale=0.2]{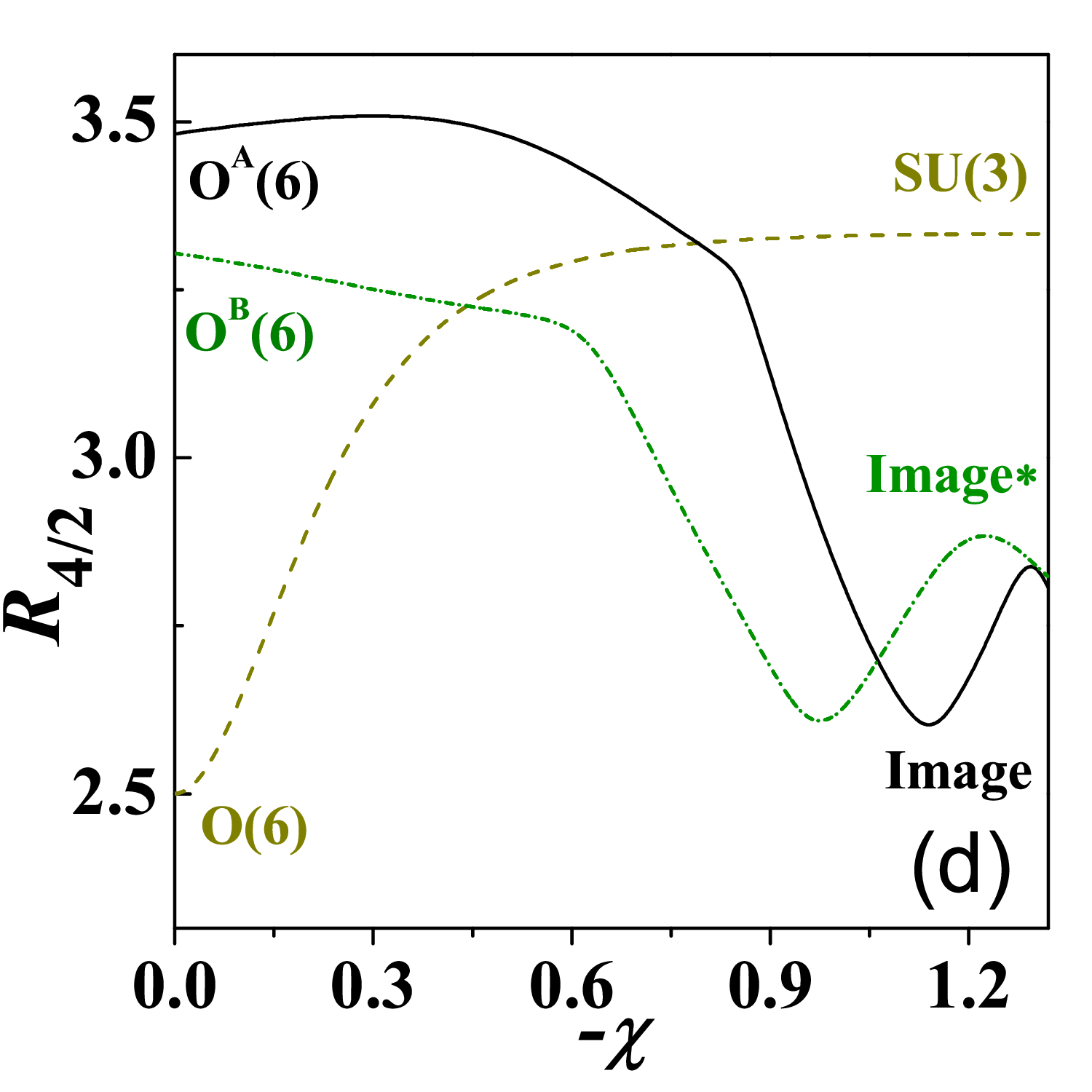}
\includegraphics[scale=0.2]{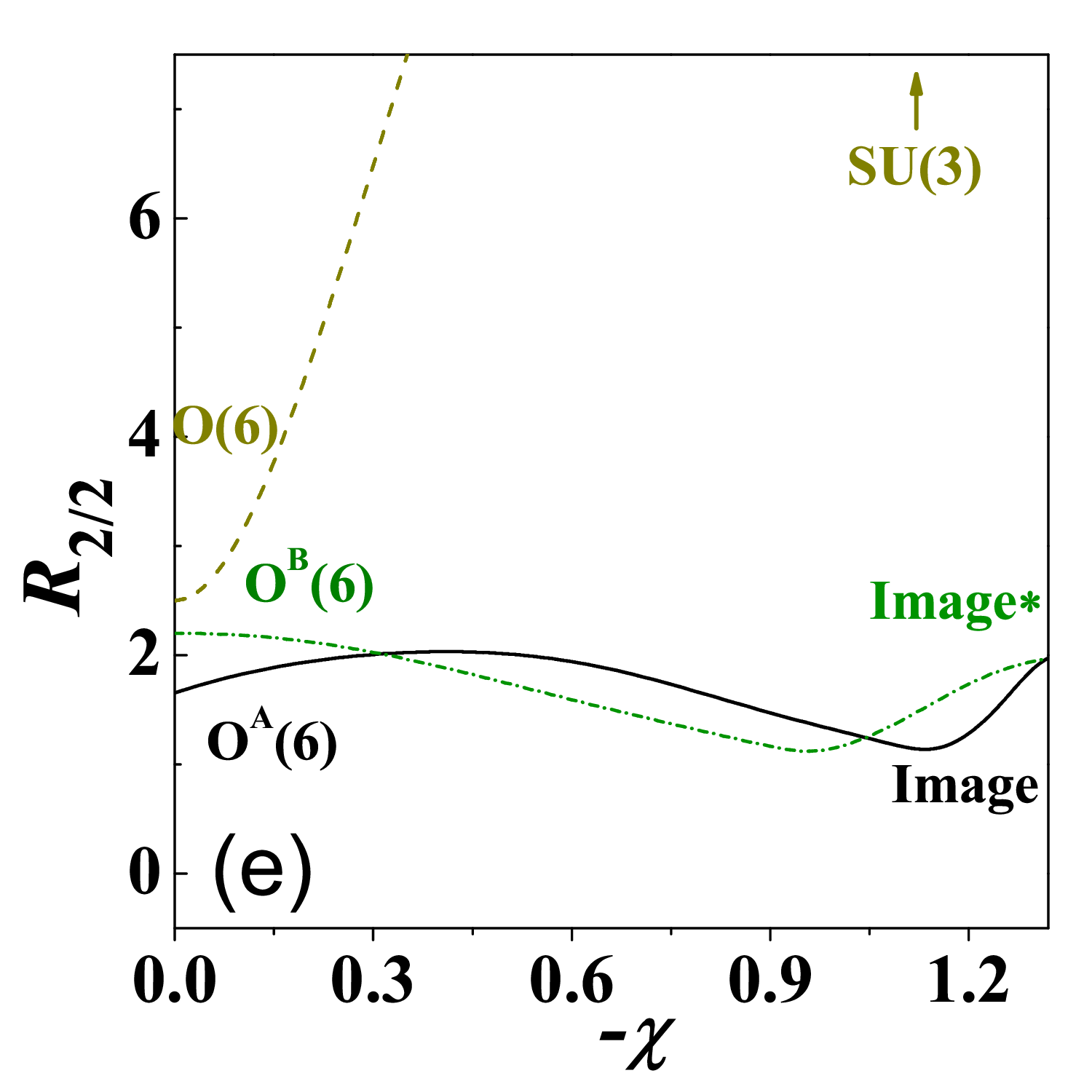}
\includegraphics[scale=0.2]{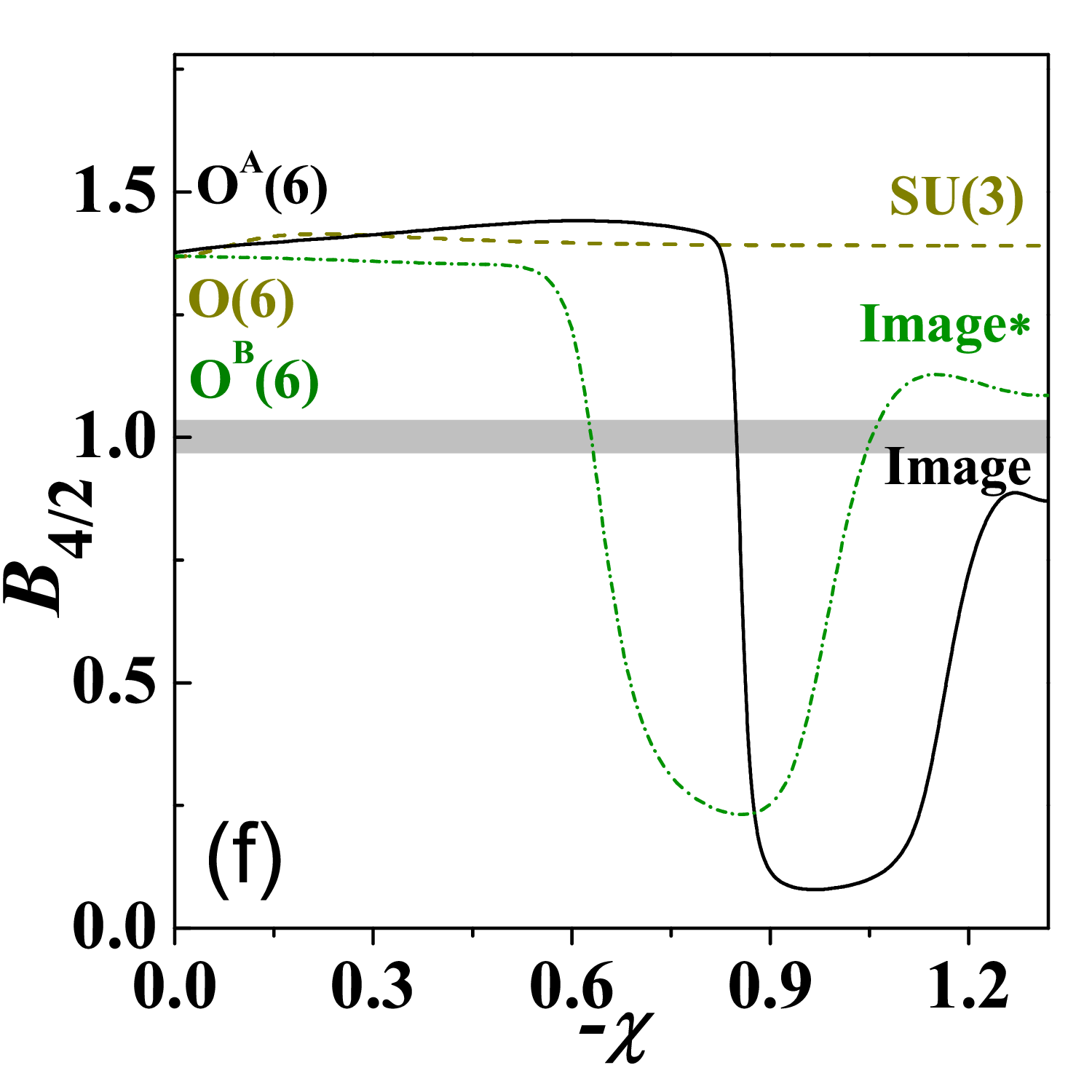}
\caption{Upper: The $R_{4/2}$ (a), $R_{2/2}$ (b) and  $B_{4/2}$ (c) values solved from (\ref{u5-rotor}) as a function of $\eta$ after fixing $\chi=-\sqrt{7}/2$ are shown to compare with those for the U(5)-SU(3) and U(5)-O(6) transitions described by the consistent-$Q$ Hamiltonian. Lower: The $R_{4/2}$ (d), $R_{2/2}$ (e) and  $B_{4/2}$ (f) values solved from (\ref{u5-rotor}) as a function of $\chi$ after fixing $\eta=1$ are shown to compare with the results for the O(6)-SU(3) transition described by the consistent-$Q$ Hamiltonian. The curves denoted by Image and Image$^{\ast}$ represent the results solved from the rotor image with $\gamma_\mathrm{~S}=30^\circ$ ($a_1:a_2:a_3=-\frac{27+10N}{3N}:1:1$) and $\gamma_\mathrm{~S}=17.8^\circ$ ($-\frac{27+10N}{3N}:1:0.9$), respectively. The total boson number adopted in the calculations is $N=9$. O(6) in each case indicates that the values are obtained for $\chi=0$ and the grey color in (c) and (f) is used to symbolize $B_{4/2}=1.0$.}\label{F5}
\end{center}
\end{figure*}

Besides the finite-$N$ corrections, another advantage of the IBM is that the competition or mixing among different collective modes can be easily handled.
To analyze the competition between the triaxial rotor and the other modes as well as its indications to the experiments, we
adopt the model Hamiltonian with
\begin{equation}\label{u5-rotor}
\hat{H}=\epsilon\,\hat{n}_d+\kappa'\,\hat{H}_\mathrm{Tri}^\chi\, ,
\end{equation}
where $\epsilon$ and $\kappa'$ are two parameters. For the convenience of the subsequent
the theoretical discussion, the parameters are reset with $\epsilon=1-\eta$ and $\kappa'=-\eta/a_1$,
in which $\eta$ is a control parameter with $\eta\in[0,~1]$ and $a_1$ is the parameter as same as that
adopted in (\ref{HS}).
In (\ref{u5-rotor}), $\hat{H}_\mathrm{Tri}^\chi$ is just the rotor mode Hamiltonian ${\hat{H}_\mathrm{Tri}}$ defined in (\ref{Tri}) except that the $\hat{Q}$ operator defined in (\ref{SQI}) has been replaced in by
$2\sqrt{2}\hat{Q}^\chi$ as defined in (\ref{Qx}).
In this way, both the $\hat{L}$ and $\hat{Q}$ operators are still the SU(3) group generators for $\chi=-\sqrt{7}/2$ but become
to be the O(6) group generators for $\chi=0$,
which means that the effect of the SU(3) symmetry breaking
on the triaxial rotor modes can be examined by varying $\chi$
with $\chi\in[-\sqrt{7}/2,~0]$.
Similarly,
the competition between the U(5) mode (spherical vibrator) and triaxial rotor can be described by the Hamiltonian (\ref{u5-rotor}) through changing $\eta$ with $\eta\in[0,~1]$ but fixing $\chi=-\sqrt{7}/2$.
To provide a parallel comparison, the
transitional situations described by the consistent-$Q$
Hamiltonian as a function of $\eta$ or $\chi$
are also considered
by resetting the parameters in (\ref{CQ})
with $\varepsilon=(1-\eta)$ and $\kappa=-2\eta$.
Then, both the U(5)-SU(3) and the
U(5)-O(6) transitional cases
can be realized through varying $\eta$ with $\eta\in[0,~1]$ but fixing $\chi=-\sqrt{7}/2$ and $\chi=0$, respectively,
while the SU(3)-O(6) transition is described by fixing $\eta=1$ but varying
$\chi$ with $\chi\in[-\sqrt{7}/2,0]$.
Note that the factor $\frac{1}{a_1}$ added in (\ref{u5-rotor})
is just used to be consistent with the parametrization
in the consistent-$Q$ formalism (\ref{CQ}).

In the following, we
focus on the typical ratios,
$R_{4/2}$, $R_{2/2}$, and $B_{4/2}$ defined previously,
since these ratios can be used not only to characterize
the evolution between different modes,
but also to signify the occurrence of the $B(E2)$ anomaly.
The calculated results as a function of $\eta$ or $\chi$ are shown in FIG.~\ref{F5}.
In the calculation for the rotor image described by $\hat{H}_\mathrm{Tri}^\chi$, we select two sets of parameters: one is taken as the
same as that adopted for FIG.~\ref{F3}(C), namely $a_1:a_2:a_3=-\frac{27+10N}{3N}:1:1$, which will generate the ground-state SU(3) IRREP $(\lambda_0,\mu_0)=(6,6)$ corresponding to $\gamma_\mathrm{~S}=30^\circ$;
another one is set by choosing $a_1:a_2:a_3=-\frac{27+10N}{3N}:1:0.9$,
which yields $(\lambda_0,\mu_0)=(10,4)$
 corresponding to $\gamma_\mathrm{~S}=17.8^\circ$.
 In contrast to the former case yielding $B_{4/2}<1.0$,
 the latter case generates $B_{4/2}>1.0$,
 which means no $B(E2)$ anomaly occurring in the latter case.

As seen from FIG.~\ref{F5}(a), the ratio $R_{4/2}$ increases with increasing of $\eta$ in all the cases except for a small fluctuation
appearing near one of the rotor images.
Nonetheless, the ratio  is
 always kept within a normal range with $2.0\leq R_{4/2}\leq3.33$.
 However, the ratio $B_{4/2}$ in the transitional case
 involving the rotor mode (denoted by Image)
 corresponding to $\gamma_\mathrm{~S}=30^\circ$
  reaches
 an unexpectedly low value with $B_{4/2}\sim0.5$
 as shown in FIG.~\ref{F5}(c).
 It means that the competition from the U(5) mode will
 enhance the $B(E2)$ anomaly feature
 emerging from the triaxial rotor mode.
 In contrast, another case involving the rotor mode
 with less triaxial deformation ($\gamma_\mathrm{~S}=17.8^\circ$) still
 keeps $B_{4/2}>1.0$ within $\eta\in[0,~1]$.
 Meanwhile,
 the ratio $R_{2/2}$ in different cases
 maintains to be small as shown in
 FIG.~\ref{F5}(b) except in the U(5)-SU(3) transition case,
 in which the ratio $R_{2/2}$ increases drastically
 with increasing of $\eta$ as expected.

An even more striking feature shown in
FIG.~\ref{F5}(d-f) is that
the $B_{4/2}$ ratio may drop
to very low value for a wide range of $\chi$
during the O(6)-SU(3) transition, while $R_{4/2}\geq2.5$ is kept.
The results indicate that the
symmetry breaking can drive a less $\gamma$-deformed
system (such as that denoted by Image*)
from $B_{4/2}>1.0$ to $B_{4/2}<1.0$, while
the ratio $R_{2/2}$ maintains to be small
except in the axial SU(3) symmetry case.
In short, small $B_{4/2}$ ratio is
relatively easier to be produced in a soft triaxial system,
which thus provides a new clue to explain the
$B(E2)$ anomaly phenomena observed
in experiment. Here, a soft triaxial system means
that deviating from the SU(3) symmetry
because the exact SU(3) realization of a triaxial rotor may yield specific $\gamma_\mathrm{~S}$ value via Eq.~(\ref{gamma})
corresponding to $\gamma$ rigid deformation.

\begin{center}
\vskip.2cm\textbf{5. Applications to Os nuclei}
\end{center}\vskip.2cm

\begin{figure}
\begin{center}
\includegraphics[scale=0.3]{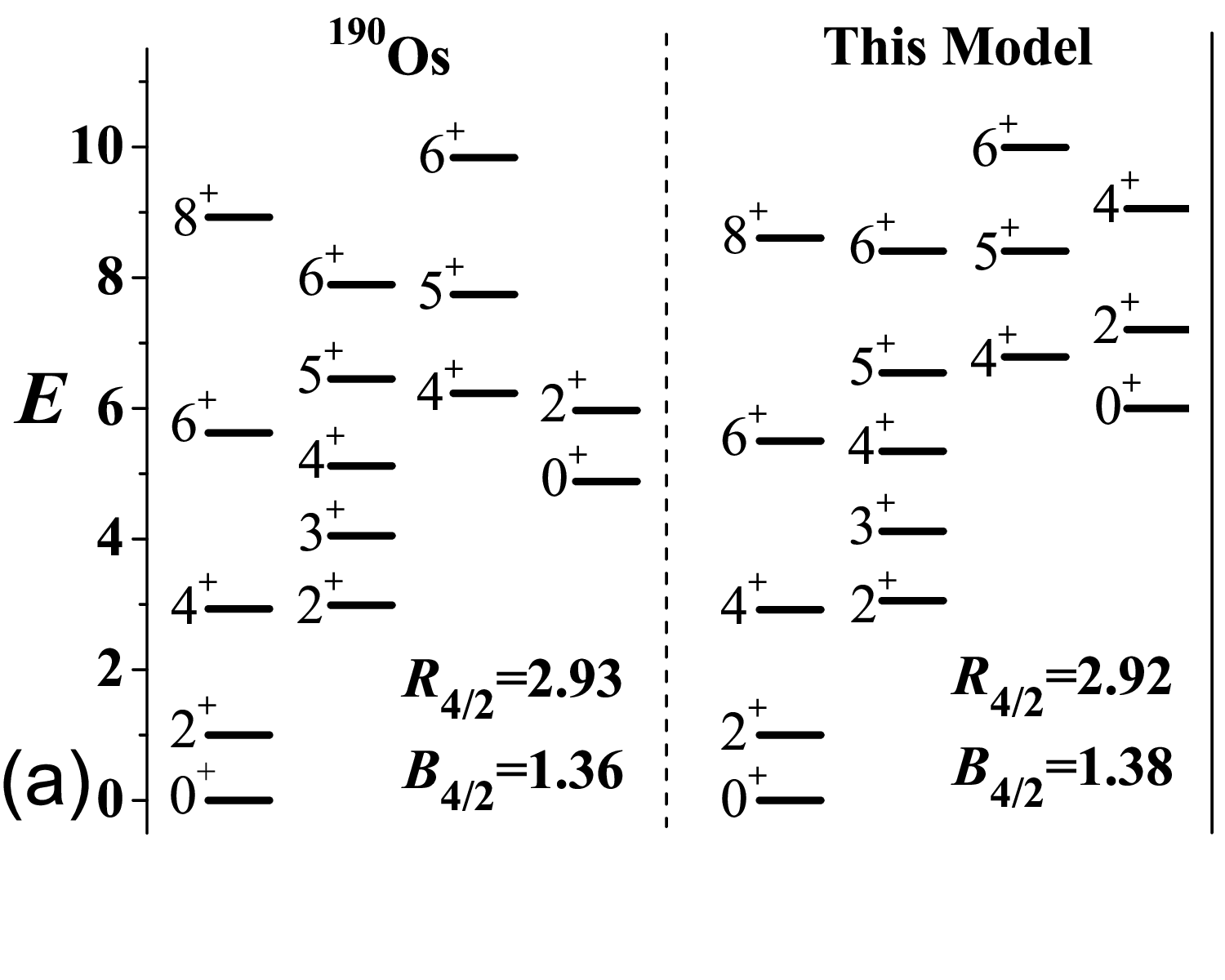}
\includegraphics[scale=0.3]{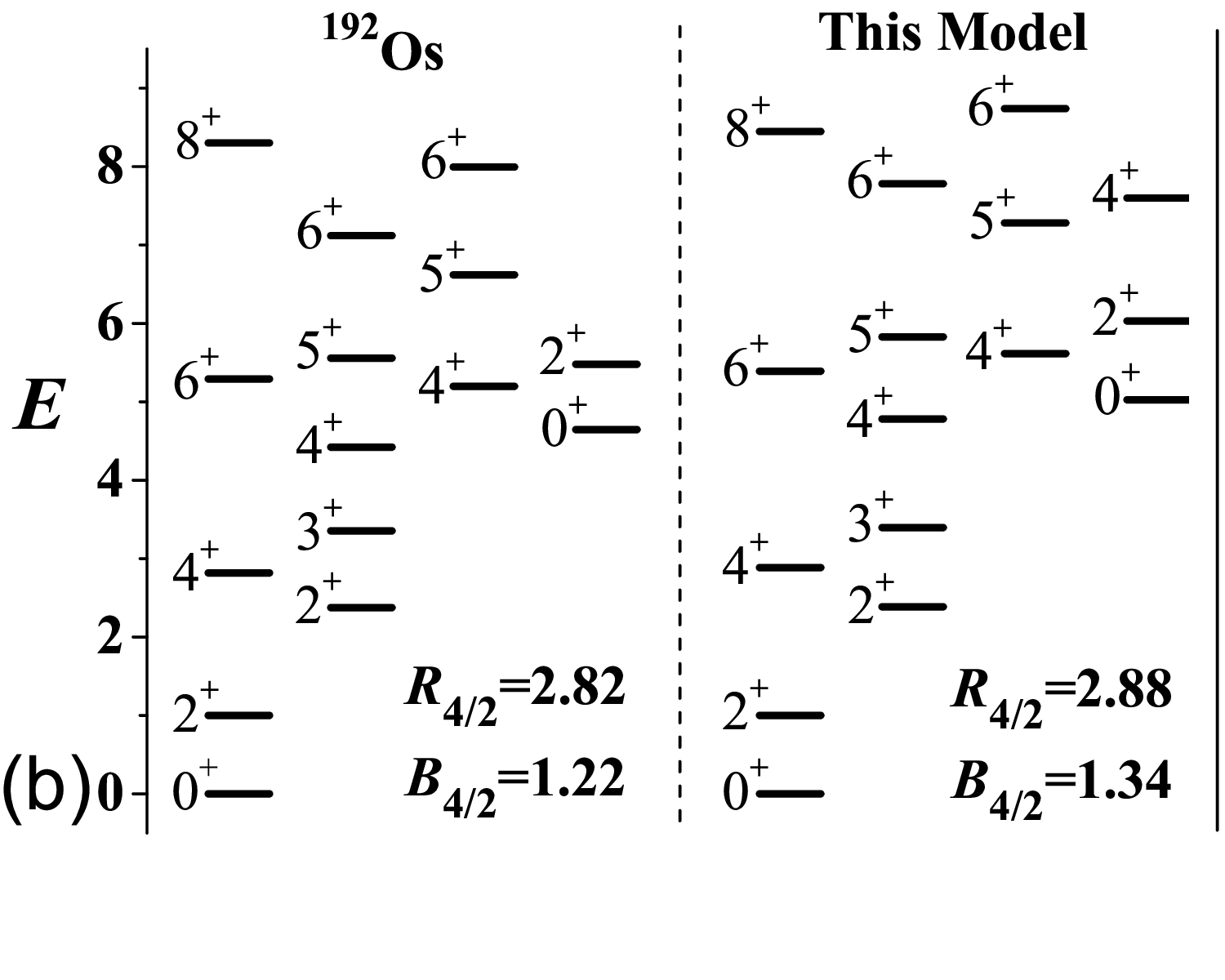}
\caption{The level patterns of $^{190,192}$Os~\cite{Singh2003,Baglin2012} with all the levels normalized to $E(2_1)=1.0$ are shown to compare with those obtained from the model Hamiltonian  (\ref{u5-rotor}) with the parameters as illustrated in the text.}\label{F6}
\end{center}
\end{figure}

\begin{table}
\caption{The model fits for the $B(E2)$ transitions (unit in W.u.) in $^{190}$Os and $^{192}$Os with the effective charges (in $\sqrt{\mathrm{W.u.}}$) adopted as $e=2.479$ and $e=2.464$, respectively. }
\begin{center}
\label{T1}
\begin{tabular}{ccc|ccc}\hline\hline
Transition&$^{190}$Os&This~Model&Transition&$^{192}$Os&This~Model\\
\hline
$2_1^+\rightarrow0_1^+$&71.9(21)&71.90&$2_1^+\rightarrow0_1^+$&62.1(7)&62.10\\
$4_1^+\rightarrow2_1^+$&105(6)&98.90&$4_1^+\rightarrow2_1^+$&75.6(20)&83.15\\
$6_1^+\rightarrow4_1^+$&113(10)&105.36&$6_1^+\rightarrow4_1^+$&100($^{+5}_{-3}$)&89.86\\
$6_1^+\rightarrow4_2^+$&6(4)&6.80&$6_1^+\rightarrow4_2^+$&$-$&$4.51$\\
$8_1^+\rightarrow6_1^+$&137(20)&99.93&$8_1^+\rightarrow6_1^+$&115(6)&82.07\\
$0_2^+\rightarrow2_1^+$&2.2(5)&0.12&$0_2^+\rightarrow2_1^+$&0.57(12)&0.30\\
$0_2^+\rightarrow2_2^+$&23(7)&57.54&$0_2^+\rightarrow2_2^+$&30.4($^{+30}_{-23}$)&45.02\\
$2_2^+\rightarrow0_1^+$&5.9(6)&7.76&$2_2^+\rightarrow0_1^+$&5.62($^{+21}_{-12}$)&5.33\\
$2_2^+\rightarrow2_1^+$&33(4)&51.00&$2_2^+\rightarrow2_1^+$&46($^{+26}_{-12}$)&58.30\\
$4_2^+\rightarrow2_1^+$&0.68(6)&0.12&$4_2^+\rightarrow2_1^+$&0.29(3)&0.08\\
$4_2^+\rightarrow4_1^+$&30(4)&38.25&$4_2^+\rightarrow4_1^+$&30.9($^{+36}_{-18}$)&32.48\\
$4_2^+\rightarrow2_2^+$&53(5)&45.14&$4_2^+\rightarrow2_2^+$&45.2($^{+14}_{-18}$)&39.17\\
$4_2^+\rightarrow3_1^+$&65(13)&38.15&$4_2^+\rightarrow3_1^+$&$-$&$11.03$\\
$6_2^+\rightarrow4_1^+$&$<$0.8&0.04&$6_2^+\rightarrow4_1^+$&$-$&$0.077$\\
$6_2^+\rightarrow4_2^+$&65(13)&59.59&$6_2^+\rightarrow4_2^+$&52($^{+3}_{-6}$)&42.60\\
$6_2^+\rightarrow6_1^+$&31(8)&24.23&$6_2^+\rightarrow6_1^+$&26($^{+55}_{-21}$)&17.78\\
$4_3^+\rightarrow2_1^+$&0.001(4)&0.02&$4_3^+\rightarrow2_1^+$&0.22($^{+21}_{-10}$)&0.17\\
$4_3^+\rightarrow4_1^+$&0.084(17)&0.001&$4_3^+\rightarrow4_1^+$&$-$&$0.081$\\
$4_3^+\rightarrow2_2^+$&7.6(15)&10.77&$4_3^+\rightarrow2_2^+$&10.6($^{+18}_{-21}$)&3.57\\
$4_3^+\rightarrow3_1^+$&27(6)&38.73&$4_3^+\rightarrow3_1^+$&56($^{+14}_{-15}$)&42.02\\
$4_3^+\rightarrow4_2^+$&14(6)&49.61&$4_3^+\rightarrow4_2^+$&24($^{+6}_{-7}$)&78.75\\
$2_3^+\rightarrow2_2^+$&$-$&$3.91$&$2_3^+\rightarrow2_2^+$&0.41(4)&3.41\\
$2_3^+\rightarrow3_1^+$&$-$&$31.25$&$2_3^+\rightarrow3_1^+$&1.1(10)&20.71\\
\hline\hline
\end{tabular}
\end{center}
\end{table}

\begin{figure}
\begin{center}
\includegraphics[scale=0.3]{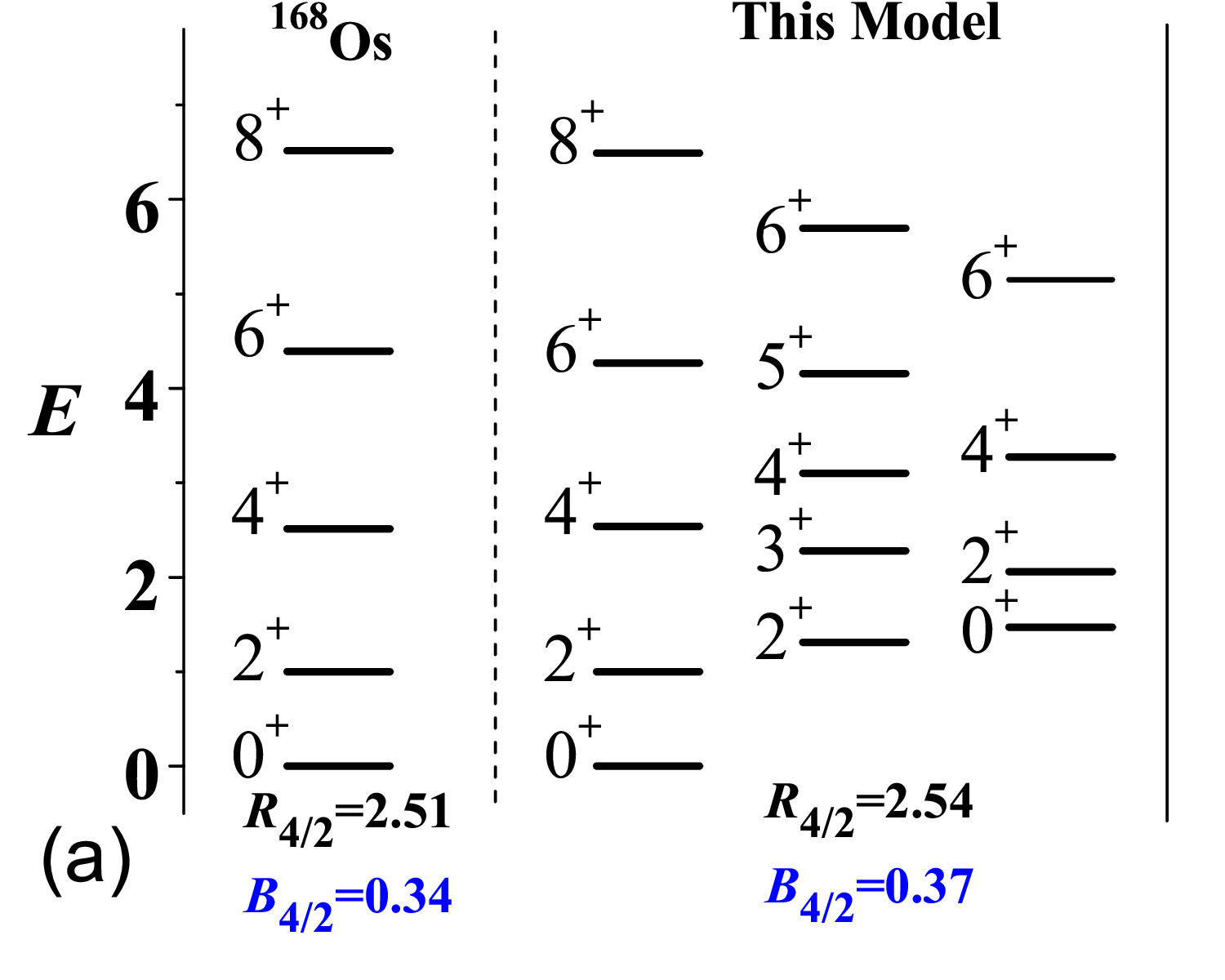}
\includegraphics[scale=0.3]{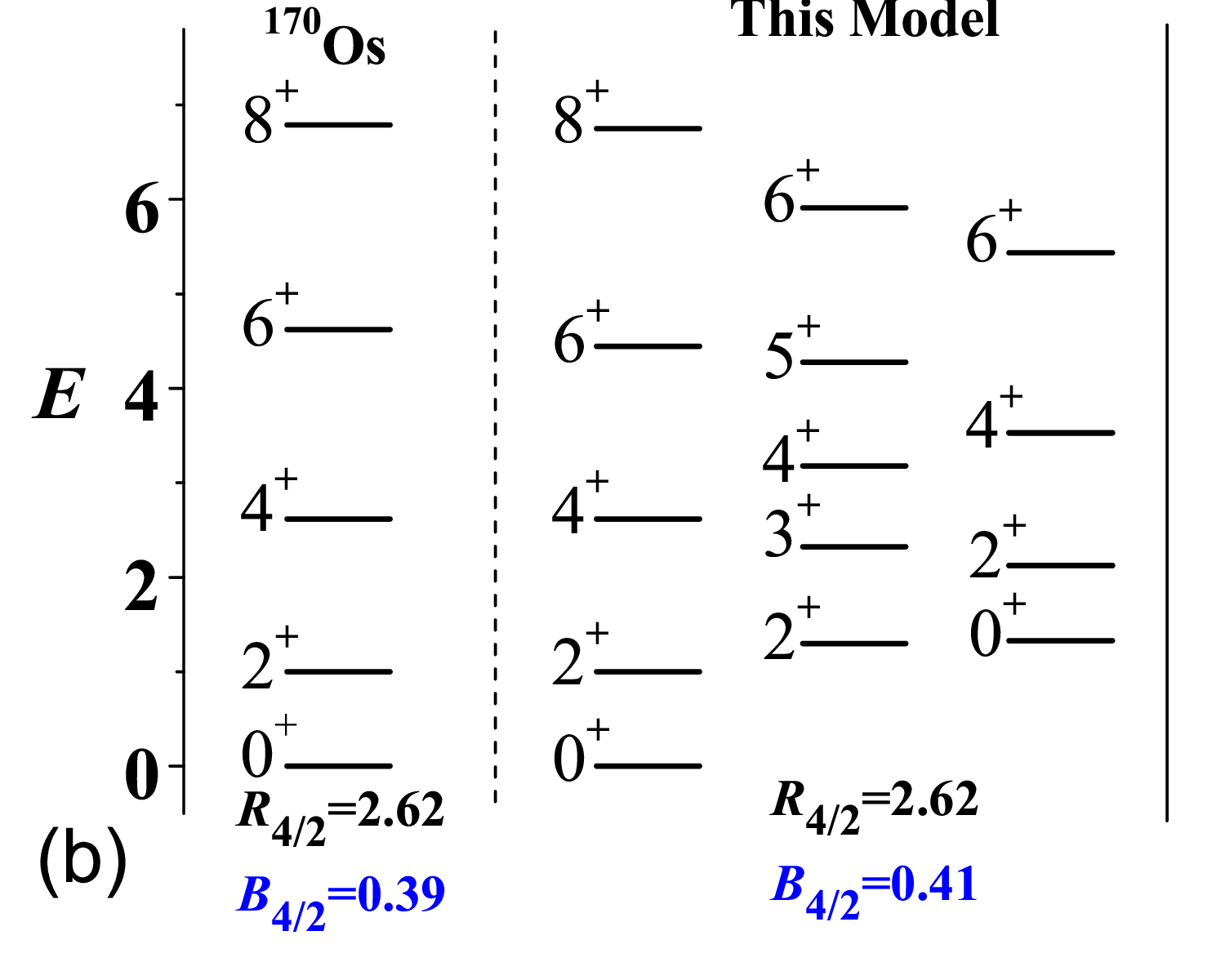}
\caption{The level patterns of $^{168,170}$Os~\cite{Grahn2016,Goasduff2019} will all the levels normalized to $E(2_1)=1.0$ are shown to those solved from the model Hamiltonian (\ref{u5-rotor}) with the parameters as mentioned in the text.}\label{F7}
\end{center}
\end{figure}

\begin{table}
\caption{The calculated $B(E2)$ values normalized to $B(E2;2_1^+\rightarrow0_1^+)=1.0$ are shown to compare with those available for $^{168}$Os~\cite{Grahn2016}
and $^{170}$Os~\cite{Goasduff2019}.}
\begin{center}
\label{T2}
\begin{tabular}{ccc|ccc}\hline\hline
Transition&$^{168}$Os&This~Model&Transition&$^{170}$Os&This~Model\\
\hline
$2_1^+\rightarrow0_1^+$&1.0&1.0&$2_1^+\rightarrow0_1^+$&1.0&1.0\\
$4_1^+\rightarrow2_1^+$&$\mathbf{0.34(18)}$&$\mathbf{0.372}$&$4_1^+\rightarrow2_1^+$&$\mathbf{0.38(11)}$&$\mathbf{0.409}$\\
$6_1^+\rightarrow4_1^+$&-&0.694&$6_1^+\rightarrow4_1^+$&-&0.709\\
$2_2^+\rightarrow0_1^+$&-&0.007&$2_2^+\rightarrow0_1^+$&-&0.008\\
$0_2^+\rightarrow2_1^+$&-&0.0001&$0_2^+\rightarrow2_1^+$&-&0.067\\
\hline\hline
\end{tabular}
\end{center}
\end{table}

As a preliminary application of the triaxial rotor description,
two neutron-rich isotopes $^{190,192}$Os are chosen to be considered
and compared with
the two neutron-deficient counterparts $^{168,170}$Os,
which possesses
the same number of bosons (the number of valence nucleon pairs)
as the former two nuclei with $N=8,~9$.
The former two Os nuclei with large $R_{4/2}$ but small $R_{2/2}$ as indicated in FIG.~\ref{F2} were suggested to be the good candidates for triaxial nuclei~\cite{Allmond2008}, while the latter two Os nuclei were observed with $R_{4/2}>2.5$ but $B_{4/2}<1.0$~\cite{Grahn2016,Goasduff2019}. Undoubtedly, the spectral features in these Os nuclei can not be explained from the traditional modes in the IBM~\cite{IachelloBook87}, which thus provides a chance to test the present theoretical scheme.
The model Hamiltonian is given by (\ref{u5-rotor}).
In experiments, the available data for the neutron-rich Os nuclei are relatively more abundant as shown
in FIG.~\ref{F6} and TABLE~\ref{T1}, where the level patterns and some $B(E2)$ values
in $^{190,192}$Os are provided
in comparison with the corresponding theoretical results.
In the model calculation, it is assumed that triaxial deformation
is built from the maximally triaxial IRREP of SU(3), which are $(\lambda_0,\mu_0)=(6,6)$ for $N=9$ and $(\lambda_0,\mu_0)=(4,6)$ for $N=8$.
Accordingly, $\gamma_\mathrm{S}=30^\circ$ and $\gamma_\mathrm{S}=35^\circ$ are yielded from the two ground-state IRREPs,
which can be generated by $\hat{H}_\mathrm{Tri}^\chi$ through setting the parameters $a_1:a_2:a_3=-4.0:1.0:1.0$ and $-4.46:1.0:1.0$, respectively.
In fact, the $\gamma$ values extracted from the present model are in good
qualitative agreement with those obtained from the proxy-SU(3) scheme~\cite{Bonatsos2017I} using the shell model IRREPs for the highest weight states as given in \cite{Bonatsos2017II},
where the results indicate $\gamma=30^\circ$ for $^{190}$Os and $\gamma=40.4^\circ$ for $^{192}$Os, respectively.
Here, we have fixed $\chi=-1.32$ in the calculation for the two neutron-rich Os nuclei. The other parameters in $\hat{H}_\mathrm{Tri}^\chi$ for $^{190}$Os ($^{192}$Os) can be evaluated via Eq.~(\ref{abc})-(\ref{lmmu}) with the inertial parameters set by $A_1:A_2:A_3=51:17:136$ ($18:16:96$), which are roughly estimated from the low-lying energies in experiments. Based on the mapping scheme,
$t_1:t_2:t_3=0.0357:0.0054:0.0001$ ($0.0432:0.0063:0.0001$) for $^{190}$Os ($^{192}$Os)
is obtained. After fixing the triaxial rotor mode based on the mapping scheme, the two parameters (in keV) in the model
Hamiltonian are
determined in fitting to
the experimental data with
$\varepsilon=327.1$ (291.1) and $\kappa'=218.1$ (264.6) for $^{190}$Os ($^{192}$Os).

Similarly, the model parameters for $^{168,170}$Os can be determined from the mapping scheme in the same way.
Due to the same boson numbers given to them, the maximal triaxial IRREPs for the two neutron-deficient nuclei are also obtained as $(\lambda_0,\mu_0)=(6,6)$ for $N=9$ and $(\lambda_0,\mu_0)=(4,6)$ for $N=8$, which means that the obtained $\gamma$ values would be as same as those for $^{190,192}$Os. However, the $\gamma$ deformations in the two neutron-deficient Os nuclei are supposed to be much softer~\cite{Grahn2016,Goasduff2019} than in the neutron-rich ones. This point in the present model will be additionally reflected from the parameter $|\chi|<1.32$ (see FIG.~\ref{F5}(f)). Specifically, the parameters for $^{168}$Os ($^{170}$Os) are adopted by $\varepsilon=66.1$ (93.2) keV, $\kappa'=82.6$ (66.6) keV together with $a_1:a_2:a_3=-4.46:1.0:1.0$ ($-4.0:1.0:1.0$) and $t_1:t_2:t_3=0.5159:0.0606:-0.0001$ ($0.5400:0.0493:-0.0003$). In addition, $\chi=-0.8$ and $\chi=-1.0$ are taken for $^{168}$Os and $^{170}$Os, respectively, corresponding to softer triaxial rotor images. By contrast, $^{168}$Os ($^{170}$Os) with ten (twelve) valence neutron particles in the proxy-SU(3) scheme has the shell model IRREP for the highest weight state being very different from that for its neutron-rich partner $^{192}$Os ($^{190}$Os) with ten (twelve) valence neutron holes, thus giving the relatively smaller triaxial deformations with $\gamma\simeq20^\circ$ ($\gamma\simeq14^\circ$).
In comparison with the neutron-rich Os nuclei, the low-lying data for the neutron-deficient Os nuclei are scarce, and only those for the yrast states in $^{168,170}$Os are provided in comparison with the model results.

It is shown in FIG.~\ref{F6}
that the level patterns of $^{190}$Os and $^{192}$Os are well
reproduced  by the model.
Especially, the lowest excitation energy in the two nuclei with
$E(2_1)=186.7$keV and $205.8$keV, respectively, is well reproduced.
As further shown in Table \ref{T1}, the consistency between experimental and theoretical results is evident, particularly in the calculated $B(E2)$ values, notably for strong intra-band transitions and weak inter-band transitions.
The only discrepancy arises from the overestimated $B(E2;2_3^+\rightarrow3_1^+)$ for $^{192}$Os, which could potentially be rectified by incorporating other modes into the model Hamiltonian.
The low-lying properties of the neutron-rich Os nuclei seem to find reasonable explanation through the mixing of triaxial rotor mode with the vibrational U(5) mode, expanding the IBM descriptions beyond the traditional modes, as depicted in FIG. \ref{F2}.
For the neutron-deficient nuclei, $^{168}$Os and $^{170}$Os, the theoretical results successfully reproduce the yrast levels and the $B(E2)$ anomaly characterized by $B_{4/2}<1.0$, as illustrated in FIG. \ref{F7}.
Additionally, theoretical calculations predict very low $\beta$ and $\gamma$ bands for these nuclei,
mirroring the neutron-rich cases as depicted in FIG. \ref{F6}. It is noteworthy from Table \ref{T2} that the $B(E2)$ anomaly, $B(E2;L_1^+\rightarrow(L-2)_1^+)/B(E2;2_1^+\rightarrow0_1^+)<1.0$, persists to the yrast states of higher spins. While a more quantitative prediction requires further experimental constraints, the current analysis strongly suggests the possibility of a soft triaxial deformation in these neutron-deficient nuclei exhibiting the $B(E2)$ anomaly, given that the model parameters are fully constrained by the mapping from the triaxial rotor. This finding is consistent with the mean-field calculations~\cite{Goasduff2019,Guzman2010}.
Moreover, collective modes with $B_{4/2}<1.0$ are observed not only in intermediate-mass nuclei~\cite{Kintish2014}, but also in light nuclei~\cite{Tobin2014}.
For instance, no-core symplectic shell model (NCSpM) calculations~\cite{Tobin2014} suggest that small SU(3) irreducible representations, such as $(\lambda,\mu)=$(4,2) and (6,2), may significantly contribute to the low-lying yrast states of $^{20}$Mg and $^{20}$O. Based on this analysis, it can be inferred that small triaxial $(\lambda,\mu)$ may induce triaxial deformation associated with $B_{4/2}<1.0$. This conclusion is supported not only by the NCSpM description~\cite{Tobin2014} but also by the possible triaxiality in $^{20}$Mg analyzed in \cite{Mitra2002}.

It is worth noting that the $B_{4/2}<1.0$ phenomenon was recently explained in \cite{Wang2020} from the prohibition of $E2$ transitions between different SU(3) irreducible representations using an IBM Hamiltonian similar to the extended consistent-$Q$ formula~\cite{Fortunato2011}, which actually allows for soft triaxial deformations in a narrow parameter region. Further extensions of the study in \cite{Wang2020} were recently made to analyze nuclear spectra related to the prolate-oblate shape transition~\cite{Wang2023I} and the associated emergent O(5)-like $\gamma$-soft modes~\cite{Wang2022,Wang2023II,Wang2024}, showing the rich SU(3) dynamics in the IBM after involving the high-order terms. In these studies, the 2nd- and 3rd-order SU(3) symmetry-conserving terms as defined in (\ref{C2})-(\ref{C3}) are particularly addressed. Since the leading SU(3) IRREPs generated by $\hat{C}_2[\mathrm{SU(3)}]$ and $\hat{C}_3[\mathrm{SU(3)}]$ are $(\lambda,\mu)=(2N,0)$ and $(0,N)$, respectively, it is not surprising that these SU(3) terms may play the central role in demonstrating the spectral evolution in the prolate-oblate shape transition~\cite{Zhang2012}. In contrast, the high-order SU(3) symmetry terms in the present scheme are introduced in a compact way via the SU(3) mapping of a quantum rotor. In particular, it is shown that the fourth-order term is necessary to generate an algebraic image of the triaxial rotor, which can not only be applied to yield a soft triaxial rotor mode associated with $B_{4/2}<1.0$ as revealed in the recent study~\cite{Zhang2022} and even earlier work~\cite{Zhang2014} (see FIG.~(2)(b)), but also generate a relatively rigid rotor mode like that occurring for $^{190,192}$Os as discussed above. Another point worth mentioning is the asymmetry between prolate and oblate in the SU(3) IBM scheme~\cite{Wang2023I,Zhang2012}, which is mainly caused by the difference between the eigenvalues of $\hat{C}_2[\mathrm{SU(3)}]$ (prolate) and $\hat{C}_3[\mathrm{SU(3)}]$ (oblate). The resulting asymmetric structural evolutions agree well with the realistic situations like the prolate-oblate shape transitions occurring in the $A=190$ mass region~\cite{Wang2023I,Zhang2012}. Nonetheless, the IBM cannot tell which nucleus is prolate, oblate or triaxial solely based on the boson number $N$. An answer to such a kind of asymmetry between prolate and oblate undoubtedly needs a microscopic model such as the proxy-SU(3) shell model scheme~\cite{Bonatsos2017I}, by which one can predict not only the locus of the prolate-oblate shape transition but also the dominance of prolate shapes in experiments~\cite{Bonatsos2017II}. It is thus highly expected to find a suitable way to extend the current IBM-based analysis into the proxy-SU(3) scheme.

\begin{center}
\vskip.2cm\textbf{6. Summary}
\end{center}\vskip.2cm

In summary, we propose a scheme for describing triaxial dynamics in finite-$N$ systems by constructing the IBM image of the rotor model Hamiltonian using the SU(3) mapping procedure. The resulting model dynamics are meticulously analyzed, revealing novel features not anticipated in traditional collective modes. Particularly noteworthy is the emergence of the $B(E2)$ anomaly feature, characterized by $R_{4/2}>2$ and $B_{4/2}<1$, which naturally arises from the IBM image of the triaxial rotor and is significantly enhanced in soft triaxial cases such as in the U(5)-SU(3) and O(6)-SU(3) transitions, as observed in FIG. \ref{F5}.
As applications, the model Hamiltonian involving triaxial rotor modes is employed to describe four Os nuclei, all presumed to be triaxially deformed. The results demonstrate that the triaxial model not only excellently reproduces the low-lying structures of the two neutron-rich Os isotopes ($^{190,192}$Os with relatively rigid triaxial deformation) but also provides a satisfactory description of the yrast levels and the suppressed $B_{4/2}$ ratios in the two neutron-deficient Os isotopes ($^{168,170}$Os with soft triaxial deformation)~\cite{Grahn2016,Goasduff2019}. This characteristic is noteworthy and merits further examination in microscopic models such as the large-scale shell and beyond mean field/generator coordinate method, which have yet to explain the anomalous $B_{4/2}$ feature.

It is notable that all typical nuclear modes, including the triaxial rotor, can be treated on equal footing within the proposed algebraic scheme. Additionally, similar $B(E2)$ anomalies in yrast states have been observed in adjacent odd Os nuclei, $^{169,171}$Os~\cite{Zhang2021}. For future research, it would be intriguing to explore the influence of odd neutrons by extending our formalism to odd-A systems.

\vskip .2cm

\begin{acknowledgments}
Support from the National Natural Science Foundation of China (12375113,11875158,12175097) and from LSU through its Sponsored Research Rebate Program as well as the LSU Foundation's Distinguished Research Professorship Program is acknowledged.
\end{acknowledgments}

\end{document}